\documentclass[journal,10pt]{IEEEtran}
\usepackage{subfigure}
\PassOptionsToPackage{subfigure}{tocloft}
\usepackage{CJK}
\usepackage{algorithm}
\usepackage{algpseudocode}
\usepackage{psfrag}
\usepackage{graphicx}
\usepackage{epstopdf}
\usepackage{multirow}
\usepackage{cite}
\usepackage{arydshln}
\usepackage{enumerate}
\usepackage{bbm}
\usepackage{bm}
\usepackage{extarrows}

\usepackage{float}
\usepackage{stfloats}

\usepackage{multirow}
\usepackage[dvipsnames]{xcolor}

\usepackage{amsthm}
\usepackage{amsmath}
\usepackage{amssymb}

\theoremstyle{plain}

\usepackage{xcolor}

\usepackage{changes}

\usepackage{etoolbox}

\usepackage{physics}

\let\mybibitem\bibitem
\renewcommand{\bibitem}[1]{%
\ifstrequal{#1}{Alex2023}{\mybibitem{#1}}
{\color{black}\mybibitem{#1}
}%
}

\usepackage{lipsum}

\title{Channel Estimation and Passive Beamforming for Pixel-based Reconfigurable Intelligent Surfaces with Non-Separable State Response}
\author{Huayan Guo, \IEEEmembership{Member, IEEE}, Junhui Rao, \IEEEmembership{Graduate Student Member, IEEE}, \\
Alex M. H. Wong, \IEEEmembership{Senior Member, IEEE}, Ross Murch, \IEEEmembership{Fellow, IEEE}, and Vincent~K.~N.~Lau, \IEEEmembership{Fellow, IEEE},
\thanks{This work was supported in part by the National Natural Science Foundation of China under Grant 62101472, in part by the Hong Kong Research Grants Council Collaborative Research Project Grant C6012-20G, and in part by the Research Grants Council under the Areas of Excellence scheme grant  AoE/E-101/23-N (Corresponding author: Vincent K. N. Lau.)}
\thanks{Huayan Guo, Junhui Rao, Ross Murch,  and Vincent K. N. Lau are with the Department of Electronic and Computer Engineering, The Hong Kong University of Science and Technology, Clear Water Bay, Kowloon, Hong Kong 999077 (e-mail: { eeguohuayan@ust.hk; jraoaa@connect.ust.hk; eermurch@ust.hk; eeknlau@ece.ust.hk}).}
\thanks{Alex M. H. Wong is with the Department of Electrical Engineering, City University of Hong Kong, Hong Kong SAR, China, and also with the State Key Laboratory of Terahertz and Millimeter Waves, City University of Hong Kong,
Hong Kong SAR, China (e-mail: {alex.mh.wong@cityu.edu.hk}).}
}

\begin{document}
\maketitle
\IEEEpeerreviewmaketitle
\begin{abstract}
Pixel-based reconfigurable intelligent surfaces (RISs) employ a novel design to achieve high reflection gain at a lower hardware cost by eliminating the phase shifters used in traditional RIS. However, this design presents challenges for channel estimation and passive beamforming due to its non-separable state response, rendering existing solutions ineffective. To address this, we first approximate the non-separable RIS response functions using a kernel-based method and a deep neural network, achieving high accuracy while reducing computational and memory complexity. Next, we propose a simplified cascaded channel model that focuses on dominated scattering paths with limited unknown parameters, along with customized algorithms to estimate short-term and long-term parameters separately. Finally, we introduce a low-complexity passive beamforming algorithm to configure the discrete RIS state vector, maximizing the achievable rate. Our simulation results demonstrate that the proposed solution significantly outperforms various baselines across a wide SNR range.
\end{abstract}

\begin{IEEEkeywords}
Intelligent reflecting surface (IRS), reconfigurable intelligent surface (RIS), non-separable state response, channel estimation, passive beamforming
\end{IEEEkeywords}

\section{Introduction}\label{sec:introduction}
Reconfigurable intelligent surfaces (RISs) are flat surfaces that alter the propagation environment by manipulating the properties of reflected waves
\cite{RIS6G2021mag,Rao2023,Chen2016,rao2022passive,Liang2022,Zhang2022,Alex2023}.
Each element of the surface can adjust the reflection responses of incoming waves independently by switching among different RIS states. Through proper configurations using passive beamforming techniques, {the} RIS can establish a virtual line-of-sight (LoS) propagation path between the transmitter and receiver. This capability offers potential benefits in various areas, including cellular communication \cite{RISBF2019huangTWC,Youli2020TSPEESERISuplink,RISBF2019guoTWC}, mobile edge computation \cite{RISedge2020jsac,RISedge2021tvt,RISedge2022tvt}, and air-ground communication \cite{RISUAV2021TWC,RISUAV2023TC,RISUAV2021JSAC}.
The RIS discussed in these papers uses phased array with phase shifters that have 2 or more bits of resolution. However, phase shifters are costly in practice and occupy space on the RIS, limiting design flexibility. To tackle this issue, a new design called pixel-based RIS has been introduced in \cite{rao2022passive}. This design allows flexible control over the radiation elements without needing phase shifters. To fully utilize its benefits, effective channel estimation and passive beamforming techniques for the pixel-based RIS are essential.


\begin{figure}
[!t]
\centering
\includegraphics[width=.95\columnwidth]{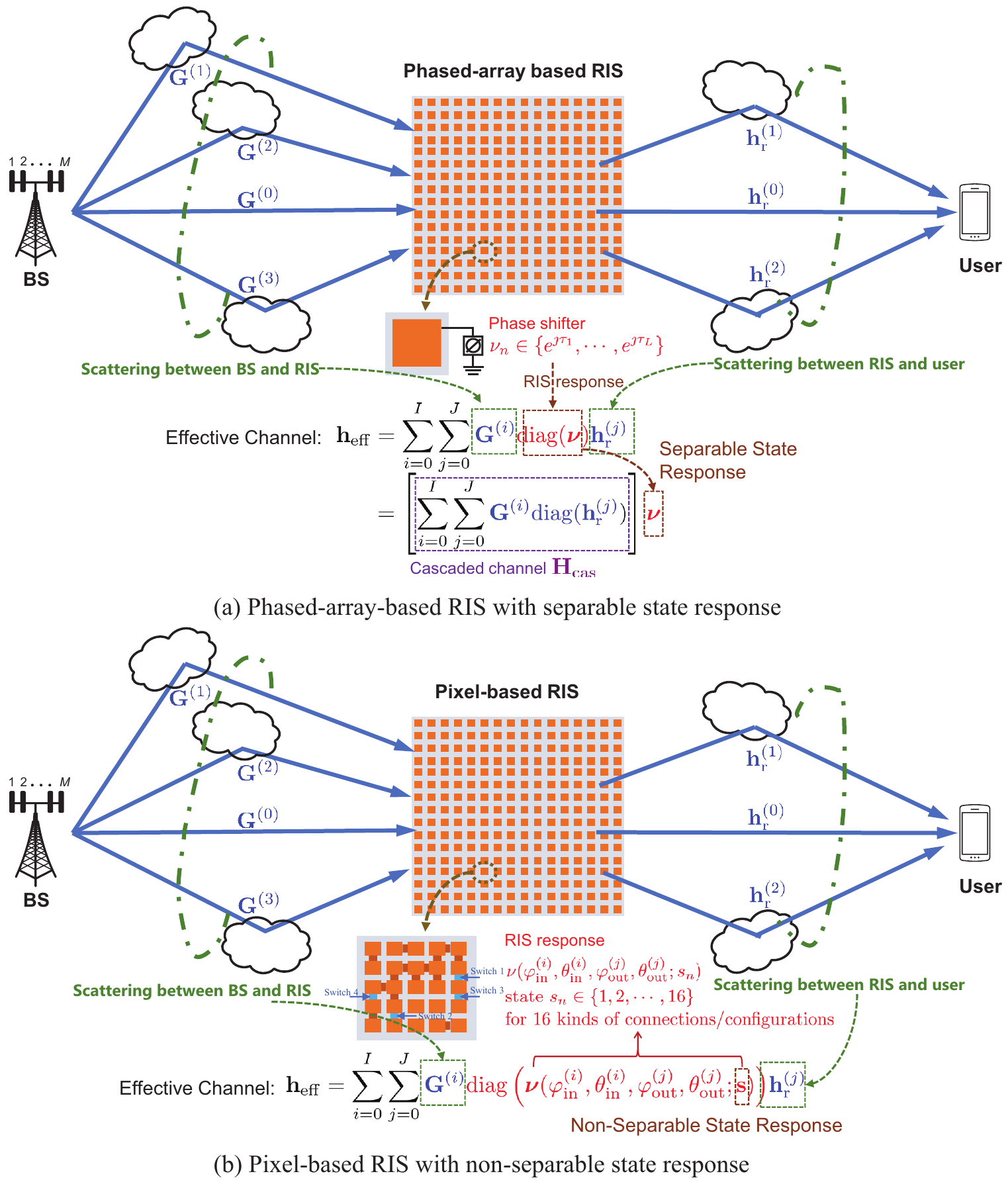}
\caption{Illustration of the RIS-assisted downlink MISO system.}
\label{IRS_illustration}
\end{figure}

For traditional phased-array RIS, designing passive beamforming usually needs accurate channel estimation of the base station (BS)-RIS-user link, as shown in {\figurename~\ref{IRS_illustration}(a)}. The effective channel between the BS and the user (${\bf h}_{\rm eff}$) can be modeled  as the multiplication of separable components, including the wireless channel from the BS to the RIS ($\bf G$),  the phase shifter network (${\bm \nu}$),  and the wireless channel from the RIS and the user (${\bf h}_{\rm r}$). Note that the phase shifter network configuration ${\bm \nu}$ can be separated out (see {\figurename~\ref{IRS_illustration}(a)}). This allows the effective channel to be represented as the product of a {\em cascaded channel response} (${\bf H}_{\rm cas}$), (which encapsulated the scattering effects in the BS-RIS and RIS-user links and is independent to the phased array setup), and a {\em RIS response term} ${\bm \nu}$ which depends only on the phased array configuration \cite{RISCE2020twc,RISCE2022Guo,RISCE2022twc,RISCE2020JSAC,RISCE2024TWC,RISCE2024TCOM,Chenjie2023RISCE}.
As a result, the conventional phased array RIS shows a {\em separable channel response}. This property allows the use of standard channel estimation algorithms \cite{MIMOLS2007CL,CEMIMOOFDM2007TSP,CEMIMO2010mag,MIMOCE2019TSP,CEmMIMO2020TWC} to estimate ${\bf H}_{\rm cas}$ from data collected with different phased array configurations. The estimated ${\bf H}_{\rm cas}$ can then be used to predict beamforming performance for other configurations, making it useful for beamforming design.


However, the pixel RIS results in a BS-RIS-user channel with a {\em non-separable state response}. Specifically, the effective BS-RIS-user channel in the  pixel-RIS-assisted system, ${\bf h}_{\rm eff}$,  is a non-separable function of the RIS configuration (i.e., states $\bf s$) and the angle of departure/arrivals ($\varphi_{\rm AoD},\theta_{\rm AoD},\varphi_{\rm AoA}, \theta_{\rm AoA}$) of the scattering paths in the BS-RIS and RIS-user links, as shown in {\figurename~\ref{IRS_illustration}(b)}. The RIS responses ${\bm \nu}$ will be different for different pairs of incident and reflection rays, and the influence of the RIS state configuration {$\bf s$}  cannot be isolated as a separable term in the effective BS-RIS-user channel.  Passive beamforming poses a challenge for RIS with a non-separable state response, as existing channel estimation and passive beamforming strategies become inapplicable. This is because the channel estimation for one RIS configuration cannot be directly  used to predict the effective channel response of the  pixel-based  RIS system at other configurations and hence, cannot be used directly for beamforming design.

Additionally, there is no simple closed-form expression for the 4D array response function for each state of the pixel RIS due to its complicated electromagnetic scattering properties. One method is to create a 4D lookup table and then use curve fitting tools to approximate the response function. However, a lookup table with 1-degree angle resolution would need 3.5 GB of memory, and storing the fitting parameters would need even more space, making this method impractical in terms of memory and time. Therefore, designing a practical approximation for the RIS response function that has low computational and memory complexity is highly desirable for both channel estimation and performance evaluation of the pixel-RIS-assisted wireless system.

Furthermore, passive beamforming in the pixel-RIS-assisted system involves  discrete optimization variables to adjust each RIS element to an appropriate state. Continuous relaxation \cite{RISdiscrete2021TCOM,RISdiscrete2022tvt,RISdiscrete2021WCL,RISdiscreteSDP2020JSAC} is employed for traditional phased array RIS, followed by regularization or projection to minimize the relaxation error by leveraging candidate phase values that are uniformly located around a unit circle. However, this property does not hold for the pixel RIS, and a new method is required to tackle the discrete state selection challenge.

More generally, these issues can arise with any RIS that uses elements that change their own structure, instead of adjusting the reflected phase through an external port connected to a phase shifter. This is because reconfiguring the structure alters the element's scattering characteristics, including the magnitude and directionality, resulting in non-separable states. For example, the pixel based elements focused on in this work fall into this category, as the switches between pixels change the element structure. Furthermore, any element made from sub-elements will potentially also fall into this category, including those  elements with switches and  elements made up of sub-elements in metasurfaces \cite{Liang2022,Zhang2022,Alex2023}. While this work focuses on the pixel based RIS further investigation on other RISs may also reveal non-separable state properties that need to be considered.

In this paper, we investigate passive beamforming and channel estimation for a multiple-user multiple-input-single-output (MU-MISO) system assisted by pixel-based RIS with a non-separable state response. The following summarizes the key contributions of this work.


\begin{itemize}
\item  {\bf {Compact Representation of RIS Response Functions for Reduced Computational and Memory Usage}}:
    Accurate and low-overhead representation of the RIS response is essential for channel estimation and passive beamforming design. To achieve this, we first propose a product-Legendre-kernel-function-based approximation, extending existing antenna radiation pattern approximation methods \cite{kernelfitting2023Gcom,kernelCE2024TC} from 2D to 4D functions. Additionally, we introduce a task-specific deep neural network (DNN) to approximate the RIS response function. This DNN employs a product structure to address features from incident and reflection angles, incorporating multiple parallel blocks that mimic kernel functions with the same structure but different weights. The kernel-based solution achieves a normalized mean square error (NMSE) of around 0.001 in most cases, while the DNN-based solution further reduces the NMSE by about 100 times with much lower computational complexity.

\item  {\bf {Effective Estimation and Prediction of Cascaded RIS Channel with Non-separable State Response}}:
To tackle the challenge of channel estimation in the RIS system with a non-separable state response, we propose a simplified channel model focusing on dominated scattering paths with limited unknown parameters. We first introduce a three-step estimation algorithm for instantaneous channel responses given incident/reflection angles. Building on this, we develop an algorithm to estimate these angles using historical channel observations, as they are long-term parameters that vary slowly. Those estimated channel responses and angles are then substituted into the proposed model for channel prediction. Simulations show that the proposed algorithm accurately predicts cascaded channels with new RIS state vectors that have not been observed during channel sounding.


\item  {\bf {Low-Complexity Passive Beamforming Algorithm for Discrete RIS State Configuration}}:
    The state configuration in the RIS system corresponds to passive beamforming, modeled as a discrete optimization problem regarding the discrete state vector. To address the combinatorial optimization challenge, we reparameterize each discrete state variable with a one-hot state selection vector, which is relaxed to a continuous-valued vector representing selection probabilities. A sparse regularization constraint is then introduced to encourage the selection probability vector to approach a one-hot configuration. Simulation results demonstrate that the proposed passive beamforming algorithm significantly outperforms various baselines.
\end{itemize}

\section{System Model}\label{sec:system_model}

\subsection{RIS with Separable and Non-Separable State Responses}
The concept of RIS with separable and non-separable state responses have been described in the introduction and also illustrated in Fig. 1. Further details on how these are commonly implemented are described next.

\subsubsection{Phase-Shifting Array-based RIS}

\begin{figure}[!t]
\begin{centering}
\textsf{\includegraphics[width=0.6\columnwidth]{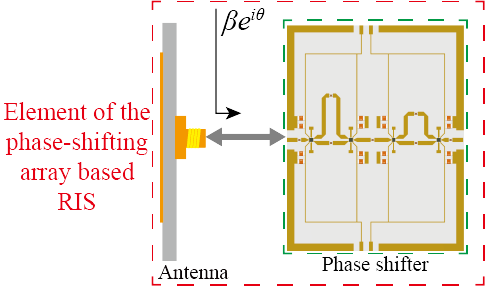}}
\par\end{centering}
\caption{The architecture of phase-shifting array based RIS element.}
\label{element of phase shifting RIS}
\end{figure}

A common implementation of RIS is the phase-shifting array-based RIS,
often referred to as a reflectarray, {which is an RIS with a separable state response.}
As shown in the Fig. \ref{element of phase shifting RIS},
each element in this type of RIS consists of an antenna loaded with
a phase shifter. The antenna receives the signal from the environment,
which is then reflected and its phase adjusted by the phase shifter
before being re-radiated into free space. This process allows the
phase of the reflected signal to be altered without affecting the
magnitude, thus separating the reflected phase from the effective
channel.

An important electromagnetic property that is often overlooked in the reflectarray approach
is that only about half of the incident signals can be controlled by
the phase shifter. The total scattered
field by the antenna can be divided into two components \cite{Harrington1964}:
the structure scattering term, which is independent of the antenna
load, and the antenna mode scattering term, which is tunable by the
phase shifter. Typically, antennas with metal grounds, such as patch
antennas, are used in RIS applications, resulting in approximately half of the incident
power being scattered as the structure scattering term \cite{Hansen1989}.
That is, only approximately half of the scattered signal can be controlled by the RIS  and consequently, phase-shifting array-based RIS inherently suffers an approximately 3 dB loss.
Furthermore, phase shifter losses can also
{be high, especially for the common configuration shown in {\figurename~\ref{element of phase shifting RIS}} where the signal passes both forward and backward through the phase shifter, leading to an additional loss of 2 dB or more.}


\subsubsection{Pixel-based RIS}

\begin{figure}[!t]
\begin{centering}
\textsf{\includegraphics[width=0.30\columnwidth]{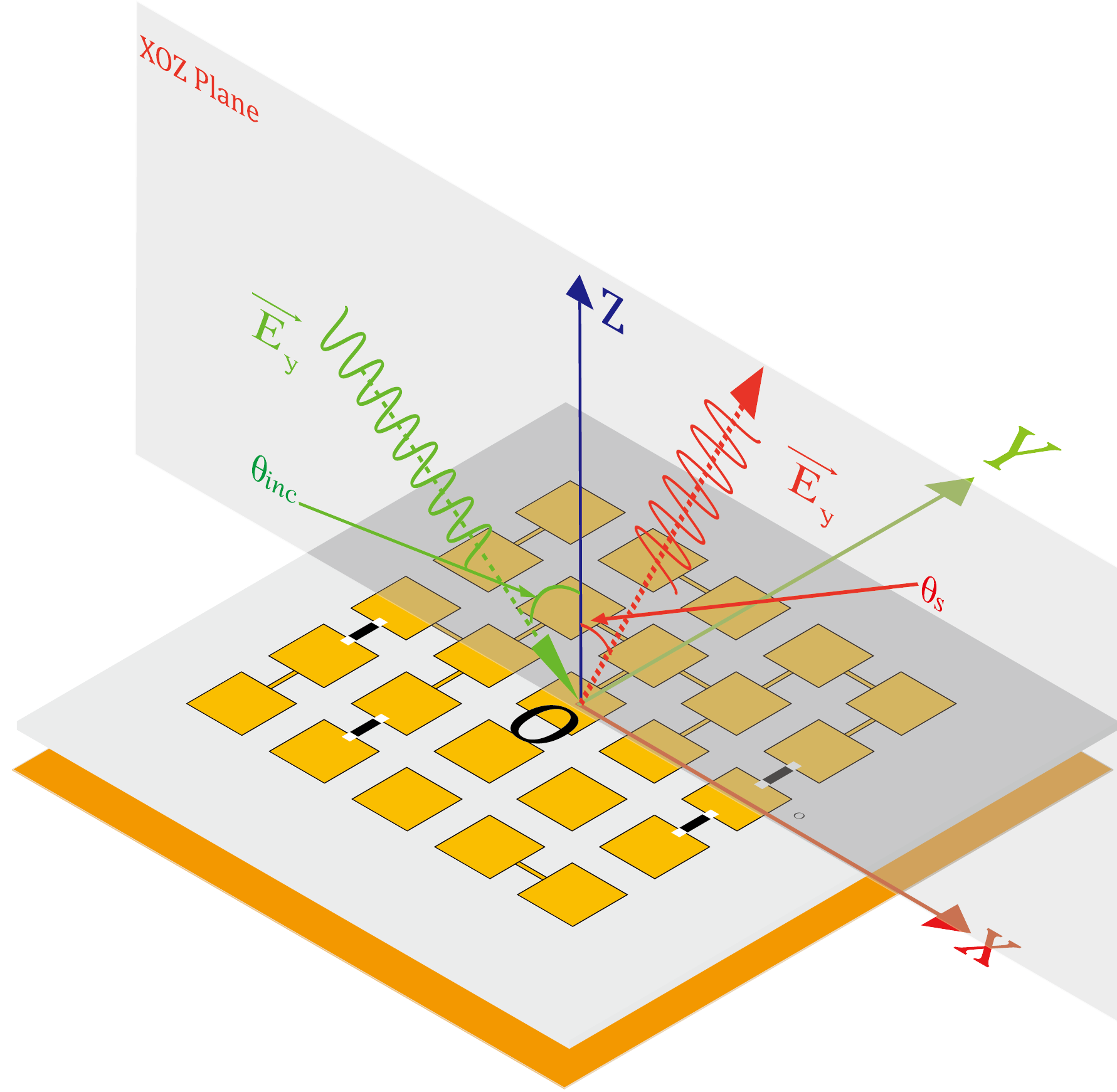}\includegraphics[width=0.55\columnwidth]{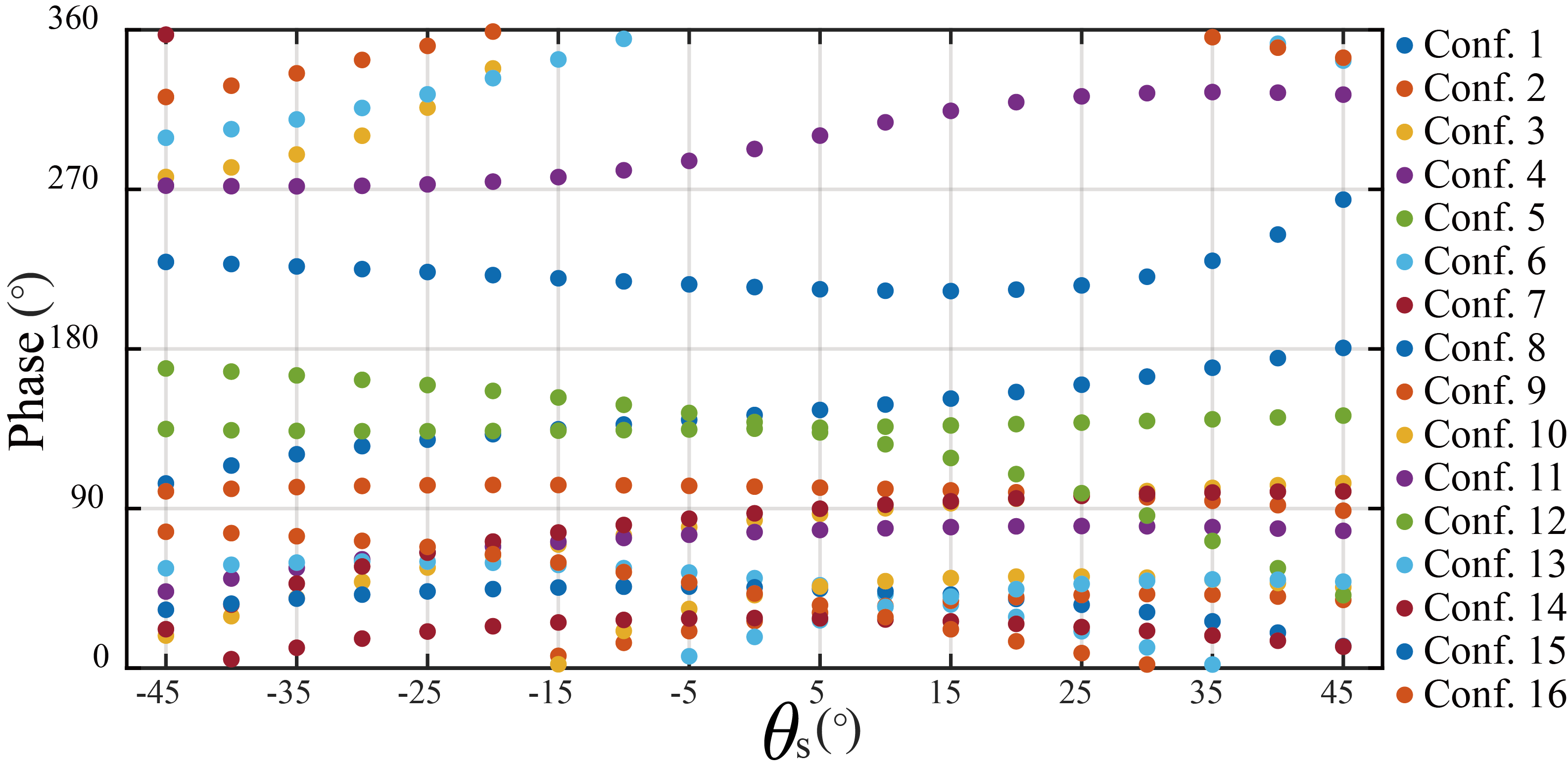}}
\par\end{centering}
\begin{centering}
(a)\hspace*{0.5\columnwidth}(b)
\par\end{centering}
\caption{(a) Details of pixel-based RIS element and coordinate system with
incident and reflected angles shown. (b) Phases of the reflected wave
are shown by colored filled circles, versus $\theta s$ {[}see (a){]}
for 16 different RIS configurations (at 2.4 GHz) when excited by vertically
incident plane wave.}
\label{element of pixel-based RIS}
\end{figure}

As shown in {Fig.~\ref{element of pixel-based RIS}(a)},
{a non-separable state response RIS is the pixel-based RIS first proposed in \cite{rao2022passive}, which}
consists of a 5$\times$5 pixel array with a metal ground
and an air gap beneath it. The dimension of each pixel is much smaller than a wavelength while the pixel array is of the scale of a wavelength. The pixels are selectively interconnected,
with four RF switches (PIN diodes) strategically placed to create
specific resonating structures at the wavelength scale. By changing
the states of these diodes, the pattern can be switched to another
with different properties. Unlike phase-shifting array-based RIS,
pixel-based RIS do not experience the 3 dB loss due to structure scattering,
as it considers the total scattered field. In addition the double losses through the external phase shifter are also avoided.
However, while the reflected
phases can be finely tuned to cover 360 degrees, they depend on both the
diode states and the angles of departure/arrival as shown in {Fig.~\ref{element of pixel-based RIS}(b)},
leading to a non-separable state response.

\subsubsection{{\color{black}Motivation for the Development of Pixel-Based RIS}}
For flexible control of RIS elements, the phased array needs phase shifters (and for full pattern control elements also ideally need to have amplitude control). However, these items are difficult to prototype  and are not cost effective when offering 2 or more bits of resolution. They also take up space on the RIS limiting design flexibility. In addition, they also increase the losses in the RIS. For instance, the signal received by an element must pass through the phase shifter twice--once for reflection and again for reradiation--resulting in a double loss effect that can increase losses by 2-3 dB per element.

In contrast, the pixel antenna approach provides a very flexible and  general formulation for controlling radiation element characteristics. Nearly any element geometry can be formed by appropriately connecting the pixels together using electronically controlled switches. This allows the full degrees of freedom within the element area to be exploited and this allows more compact designs compared to phased arrays. The pixels are connected by switches integrated into each element and are therefore compact to include. Pixel-based elements also enable fine control of the radiation characteristics and can provide a wide range of phase shifts, as shown in the published pixel RIS design and prototype \cite{rao2022passive}. Furthermore, since there are no phase shifters and signals reflect directly off the pixel surface, losses are minimized, eliminating the double loss effect.

In summary, the pixel approach to RIS design does not rely on phase shifters, resulting in reduced losses. With only a few switches needed per element, this method is cost-effective and allows for highly reconfigurable elements that can exhibit a wide variety of characteristics as required. Therefore, the pixel approach offers a competitive alternative to traditional RIS designs that utilize phase shifters.

%
%

%
%

\subsection{Scattering Model for BS-RIS and RIS-UE Channels for the Pixel-based RIS}\label{sec:system_model_scattering}
Let $M$ be the number of BS antennas, $N$ the number of RIS elements, $L$ the number of candidate RIS states, and $K$ the number of users.
We consider a propagation environment with limited scatterers with low user mobility. The wireless propagation parameters, such as scattering cluster power, path loss, and arrival/departure angles, remain constant over an extended period, while the small-scale channel coefficients for different signal paths vary across time blocks.
Specifically, according to the 3GPP channel model \cite[Section 7]{38901}, the channel from BS to RIS in the $t$-th time block is given by
\begin{align}
{\bf G}_t & = {\bf G}^{(0)}+\sum_{i=1}^I {\bf G}_{t}^{(i)},
\end{align}
where ${\bf G}^{(0)}$ is the LoS component and ${\bf G}_{t}^{(i)}$ is the NLoS component for the $i$-th scattering cluster. Define $\phi_{\rm BS}=\{\varphi_{\rm BS},\theta_{\rm BS}\}$ and $\phi_{\rm in}=\{\varphi_{\rm in},\theta_{\rm in}\}$, where $\varphi_{\rm BS}$ and $\theta_{\rm BS}$ denote the azimuth and elevation angles departing from BS to RIS, $\varphi_{\rm in}$ and $\theta_{\rm in}$ denote the incident azimuth and elevation angles to the RIS. The components in ${\bf G}_t$ are given by
\begin{subequations}
\begin{align}
{\bf G}^{(0)} & = {\bm \alpha}_{\rm BS}(\phi_{\rm BS}^{(0)}) {\bm \alpha}^{\rm H}_{\rm RIS}(\phi_{\rm in}^{(0)}), \label{equ:los_G}\\
{\bf G}_{t}^{(i)} & = a_{t}^{(i)} \sum_{d=1}^{D_i} {\bf G}_{t}^{(i,d)},\label{equ:nlos_G}\\
{\bf G}_{t}^{(i,d)} & =  e^{\jmath \varepsilon_{t}^{(i,d)}} {\bm \alpha}_{\rm BS}(\phi_{\rm BS}^{(i,d)}) {\bm \alpha}^{\rm H}_{\rm RIS}(\phi_{\rm in}^{(i,d)}), \label{equ:nlos_G_con}
\end{align}
\end{subequations}
where  ${\bf G}_{t}^{(i,d)}$ indicates the $d$-th ray in the $i$-th scattering cluster for the NLoS link and $a_{t}^{(i)} \in {\mathbb R}$ denotes the path gain, ${\bm \alpha}_{\rm BS} \in {\mathbb C}^{M \times 1}$ and ${\bm \alpha}_{\rm RIS} \in {\mathbb C}^{N \times 1}$ are the steering vectors of BS and RIS arrays, respectively,  and $\varepsilon$ denotes the random phase for different rays.

Similarly, the channel from RIS to the $k$-th user in the $t$-th time block is given by
\begin{align}
{\bf h}_{{\rm r},k,t} & = {\bf h}_{{\rm r},k,t}^{(0)}+\sum_{j=1}^J {\bf h}_{{\rm r},k,t}^{(j)},
\end{align}
where ${\bf h}_{{\rm r},k}^{(0)}$ is the LoS component and ${\bf h}_{{\rm r},k,t}^{(j)}$ is the NLoS component for the $j$-th scattering cluster of user $k$. Define $\phi_{{\rm out},k}=\{\varphi_{{\rm out},k},\theta_{{\rm out},k}\}$ where $\varphi_{{\rm out},k}$ and $\theta_{{\rm out},k}$ denote the reflection azimuth and elevation angles from RIS to the $k$-th user. The components in ${\bf h}_{{\rm r},k,t}$ are given by
\begin{subequations}
\begin{align}
{\bf h}_{{\rm r},k,t}^{(0)} & = b_{k,t}^{(0)}  {\bm \alpha}_{\rm RIS}(\phi_{{\rm out},k}^{(0)}), \label{equ:los_hr_k}\\
{\bf h}_{{\rm r},k,t}^{(j)} & = b_{k,t}^{(j)} \sum_{d=1}^{D_i} {\bf h}_{{\rm r},k,t}^{(j,d)},\label{equ:nlos_hr_k}\\
{\bf h}_{{\rm r},k,t}^{(j,d)} & =  e^{\jmath \epsilon_{k,t}^{(j,d)}} {\bm \alpha}_{\rm RIS}(\phi_{{\rm out},k}^{(j,d)}),
\end{align}
\end{subequations}
where ${\bf h}_{{\rm r},k,t}^{(j,d)}$ is the $d$-th ray in the $j$-th scattering cluster of the NLoS link and $b_{k,t}^{(j)} \in {\mathbb R}$ is the path gain,
 and $\epsilon$ denotes the random phase response.  

Let $\nu(\phi_{\rm in}^{(i)},\phi_{\rm out}^{(j)};\ell)$ denote the response of a single element in the pixel-based RIS for the $i$-th incident ray and the $j$-th reflected ray, when the element is configured to the $\ell$-th state. Let ${\bf s}=[s_{1}, s_{2}, \cdots, s_{N}]^{\rm T}$ denote the RIS state configuration vector for all the $N$ elements. The RIS response vector is given by:
\begin{equation}\label{equ:channel_RIS_resposne}
\begin{aligned}[b]
&{\bm \nu}(\phi_{\rm in}^{(i)},\phi_{\rm out}^{(j)};{\bf s})
=\Big[ \nu(\phi_{\rm in}^{(i)},\phi_{\rm out}^{(j)};\ell=s_{1})\\
& \quad \quad\quad\quad\quad\quad\quad \cdots,
\nu(\phi_{\rm in}^{(i)},\phi_{\rm out}^{(j)};\ell=s_{N}) \Big]^{\rm T}.
\end{aligned}
\end{equation}
Finally, the cascaded channel from BS to the $k$-th user at $t$-th time block becomes a function of RIS state ${\bf s}_t$:
\begin{equation}\label{equ:channel_cascaded}
\begin{aligned}[b]
{\bf h}_{k,t}({\bf s}_t)
=&\sum_{i,j} {\bf G}_{t}^{(i)} {\rm diag} \left({\bm \nu}(\phi_{\rm in}^{(i)},\phi_{{\rm out},k}^{(j)};{\bf s}_t)\right) {\bf h}_{{\rm r},k,t}^{(j)}.
\end{aligned}
\end{equation}
From \eqref{equ:channel_cascaded}, we see that to predict ${\bf h}_{k,t}({\bf s})$ for any state vector ${\bf s}$, one needs the angles $\phi_{\rm in}^{(i)}$ and $\phi_{{\rm out},k}^{(j)}$ for all propagation rays, as well as the RIS response function $\nu(\phi_{\rm in}^{(i)},\phi_{\rm out}^{(j)};\ell)$ for any states. Note that the RIS response function is determined solely by the structure of the pixel RIS element and is independent of the wireless propagation variables ${\bf G}_{t}^{(i)}$, ${\bf h}_{{\rm r},k,t}^{(j)}$, $\phi_{\rm in}^{(i)}$ and $\phi_{{\rm out},k}^{(j)}$.

\begin{figure}
[!t]
\centering
\includegraphics[width=.7\columnwidth]{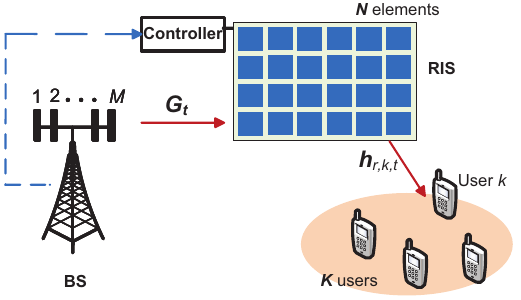}
\caption{Illustration of the pixel-based-RIS-assisted MU-MISO system.}
\label{MU_MISO}
\end{figure}

\subsection{Overall Signal Model for the Pixel-based-RIS-Assisted MU-MISO System}\label{sec:model_MU_MISO}
The overall structure of the  multiuser-(MU)-MISO communication system assisted by a pixel-based RIS is illustrated in {\figurename~\ref{MU_MISO}}. We assume that the direct link between BS and users are blocked by obstacles and hence, the signals of direct link are ignored for simplicity. In the $t$-th time block, the signal received by the $k$-th user is given by
\begin{equation}\label{equ:signal_user_k}
\begin{aligned}[b]
y_{k,t}&={\bf h}_{k,t}({\bf s}_t)^{\rm T} \sum_{k=1}^{K} {\bf w}_{k,t} x_{k,t}+ z_{k,t},
\end{aligned}
\end{equation}
where ${\bf w}_{k,t}$ denotes the precoder for the $k$-th user, $x_{k,t}$ denotes the signal of the $k$-the user, and $z_{k,t} \sim {\cal{CN}}(0,\sigma_{\rm z}^2) $ denotes the additive white Gaussian noise (AWGN) at the $k$-th user's receiver.

To simplify the analysis and focus on the passive beamforming, we utilize a regular zero-forcing (RZF) precoder. Letting ${\bf W}_t=[{\bf w}_{1,t},{\bf w}_{2,t},\cdots,{\bf w}_{K,t}]$, and
${\bf H}_{t}({\bf s}_t)=[{\bf h}_{1,t}({\bf s}_t),{\bf h}_{2,t}({\bf s}_t),\cdots,{\bf h}_{K,t}({\bf s}_t)]$.
The precoder is given by
\begin{equation}\label{equ:precoder_rzf}
\begin{aligned}[b]
{\bf W}_{t}({\bf s}_t)&= \sqrt{\gamma({\bf s}_t)} {\bf H}_t^\star({\bf s}_t) \left({\bf H}_t^{\rm T}({\bf s}_t) {\bf H}_t^\star({\bf s}_t) + \varsigma {\bf I}_K  \right)^{-1},
\end{aligned}
\end{equation}
where $\varsigma \geq 0$ is a small constant that makes the matrix inverse possible, and $\gamma({\bf s}_t)$ is used to satisfy the transmit power constraint.
The achievable rate of user $k$  is given by
\begin{equation}\label{equ:downlink_rate}
{R}_{k,t}({\bf s}_t)=  \log\left( 1+ \frac{\left|{\bf h}_{k,t}({\bf s}_t)^{\rm T} {\bf w}_{k,t}({\bf s}_t)  \right|^2 }
{\sum_{i \neq k} \left|{\bf h}_{k,t}({\bf s}_t)^{\rm T} {\bf w}_{i,t}({\bf s}_t)  \right|^2+\sigma_{\rm z}^2} \right)
.
\end{equation}
The primary objective of this paper is to configure an appropriate  RIS state ${\bf s}_t$ to maximize the sum rate: 
\begin{subequations}
\begin{align}
    {\mathcal{P}}{(\text{A})}
  \;\;\; {\bf{s}}_t^\ast  & =\; \underset{\bf{s}_t }{\rm{argmax}}  \; \;
\sum_{K=1}^{K} {R}_{k,t}({\bf s}_t) \notag\\
 & \;\; {\text{s.t.}} \;  s_{n,t} \in \{1,2,\cdots,L\}, \; \forall  n, \\
 & \quad \quad \frac{1}{K}\sum_k || {\bf w}_{k,t}({\bf s}_t)||_{\rm 2}^2 \leq P_{\rm T}.
\end{align}
\end{subequations}

\section{Approximation for the Pixel-based RIS  Response Function}
{\color{black}The pixel RIS prototype in \cite{rao2022passive} operates at a center frequency of 2.4 GHz, offering a consistent response over about 100 MHz of bandwidth (see \cite[Table II]{rao2022passive}), and its response function is independent of the wireless propagation parameters. In this section, we describe the design of approximate functions for the pixel RIS response ($\nu(\phi_{\rm in},\phi_{\rm out};\ell)$) for a single RIS element configured in the $\ell$-th state. We first propose an approximation by creating a kernel from the product of Legendre functions, followed by a DNN solution that has lower computational complexity but higher approximation accuracy.}


\begin{figure}
[!t]
\centering
\includegraphics[width=0.9\columnwidth]{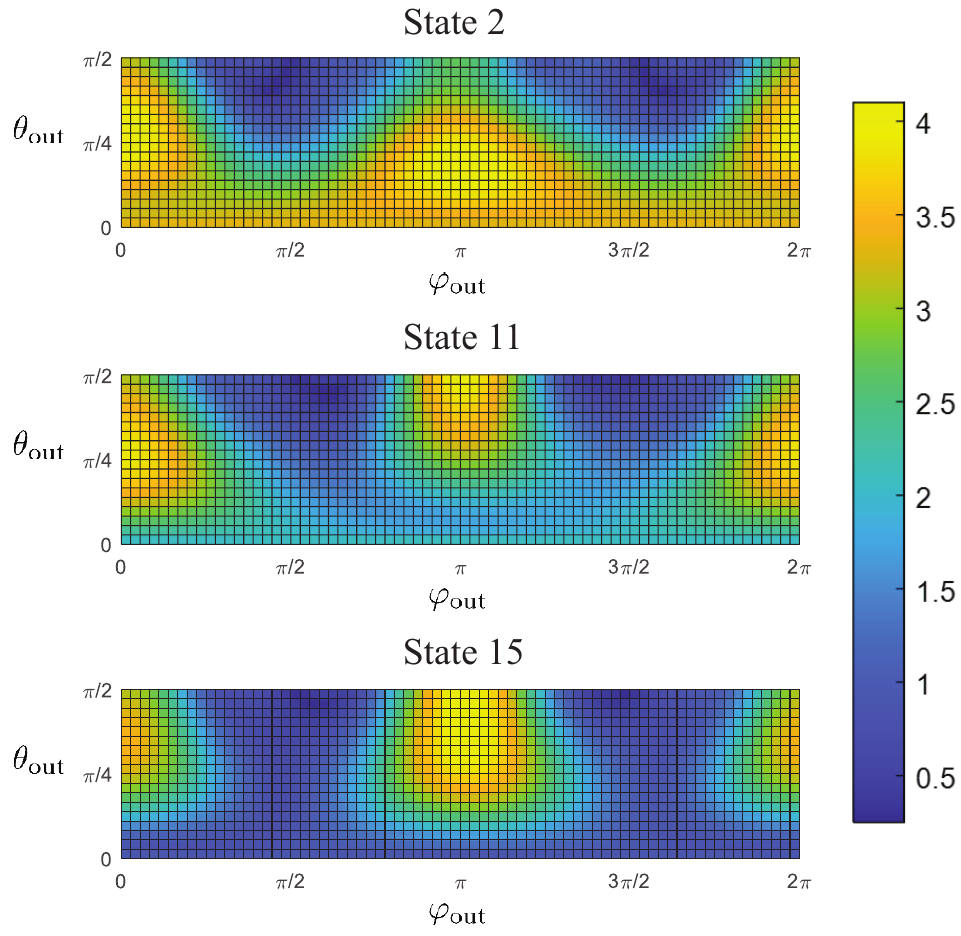}
\caption{Illustration of the magnitudes of $\nu$, for a single RIS element, for different $\varphi_{\rm out}$ and $\theta_{\rm out}$ by setting $\varphi_{\rm in}=0$, $\theta_{\rm in}=\frac{\pi}{4}$ for states 2, 11 and 15.}
\label{resp_state_1}
\end{figure}

\subsection{Product-Legendre-Kernel-Based Approximation}
The magnitudes of $\nu(\phi_{\rm in},\phi_{\rm out};\ell)$ for a single RIS element are illustrated in {\figurename~\ref{resp_state_1}} by setting $\varphi_{\rm in}=0$ and $\theta_{\rm in}=\frac{\pi}{4}$  for $\ell=1$ states 2, 11 and 15. It is seen that the response function is smooth, suggesting that it can be approximated by a weighted sum of a group of kernel functions.

While the approximation of the 4-D RIS response has not been explored, some studies have investigated the approximation of a 2-D antenna radiation pattern using Legendre-based kernel functions \cite{kernelfitting2023Gcom,kernelCE2024TC}, demonstrating good performance. {\color{black}In this paper, we generalize the 2-D Legendre-based kernel to a 4-D product-Legendre-based kernel for approximating the RIS response. The approximate function, denoted by $\overline{\nu}_\ell$, is defined as follows:
\begin{equation}\label{equ:app_fun_kernel}
\begin{aligned}[b]
\overline{\nu}_\ell( \phi_{\rm in},\phi_{\rm out}; {\bm \omega}_\ell)
&= \sum_{c_{\rm i},c_{\rm o}} \omega_{\ell, c_{\rm i},c_{\rm o}} f_{{\rm L},c_{\rm i}} (\phi_{\rm in}) f_{{\rm L},c_{\rm o}} (\phi_{\rm out}),
\end{aligned}
\end{equation}
where ${\bm \omega}_\ell=\{\omega_{\ell, c_{\rm i},c_{\rm o}}, \forall c_{\rm i}, c_{\rm o} \}$ is the parameter set to be trained, and the proposed kernel $f_{{\rm L},c_{\rm i}} (\psi_{\rm in})f_{{\rm L},c_{\rm o}} (\psi_{\rm out})$ are the product of two Legendre-based kernel functions. This product structure arises from the idea that the RIS response can be viewed as a product of the incident and reflection responses.}
Specifically, the Legendre-based kernel $f_{{\rm L},c}(\phi)$ is given by
\begin{equation}\label{equ:Legendre_kernel}
\begin{aligned}[b]
&f_{{\rm L},c}(\phi)= Y_{b}^r(\varphi,\theta)\\
&=\begin{cases}
(-1)^r \sqrt{2} \kappa_b^r \cos(r \varphi) f_{{\rm{Leg}},b}^r(\cos(\theta)), \; 0<r\leq b, \\
(-1)^r \sqrt{2} \kappa_b^r \cos(r \varphi) f_{{\rm{Leg}},b}^{-r}(\cos(\theta)), \; -b \leq r<0, \\
\kappa_b^0 f_{{\rm Leg},c}^r(\cos(\theta)), \; r=0,
\end{cases}
\end{aligned}
\end{equation}
where $c=b^2+b+r+1$, $0 \leq b \leq B$, $\kappa_b^r=\sqrt{\frac{2b+1}{4\pi} \frac{ (b-|r|)! }{ (b+|r|)! }}$, and  $f_{{\rm{Leg}},b}^r$ represents the Legendre function of degree $b$ and order $r$.

\subsection{DNN-Based Approximation}
The proposed product-Legendre-kernel-based approximation requires a relatively large $B$ to achieve high accuracy. However, the number of kernels increases with the square of $B$, leading to higher computational complexity. In this subsection, we introduce a DNN-based approximation that relies solely on linear matrix operations and simple activation functions.

\subsubsection{Data Pre-Processing}\label{sec:preprocessing}
Inspired by the Legendre-based kernel in \eqref{equ:Legendre_kernel}, we introduce new variables $\psi_{\rm in} \in {\mathbb{R}}^{3}$ and $\psi_{\rm out} \in {\mathbb{R}}^{3}$ defined as follows:
\begin{align}
\psi_{\rm in} &= [\cos(\varphi_{\rm in}), \sin(\varphi_{\rm in}), \cos(\theta_{\rm in})],\\
\psi_{\rm out} &= [\cos(\varphi_{\rm out}), \sin(\varphi_{\rm out}), \cos(\theta_{\rm out})].
\end{align}
These variables better capture the periodic nature of the azimuth angles, particularly for the responses of states 11 and 15 shown in {\figurename~\ref{resp_state_1}}.

Same as \eqref{equ:app_fun_kernel}, we denote the approximate function by $\overline{\nu}_\ell$, which becomes after replacing $\{\phi_{\rm in},\phi_{\rm out}\}$ by $\{\psi_{\rm in},\psi_{\rm out}\}$:
\begin{equation}\label{equ:app_fun}
\begin{aligned}[b]
\overline{\nu}_\ell( \phi_{\rm in},\phi_{\rm out}; {\bm \omega}_\ell)
=\overline{\nu}_\ell(\psi_{\rm in},\psi_{\rm out}; {\bm \omega}_\ell),
\end{aligned}
\end{equation}
where ${\bm \omega}_\ell$ denotes the DNN parameters to be trained.
The remaining task is to design the structure of the DNN function.

\subsubsection{DNN Structure}
\begin{figure*}
[!t]
\centering
\includegraphics[width=1.95\columnwidth]{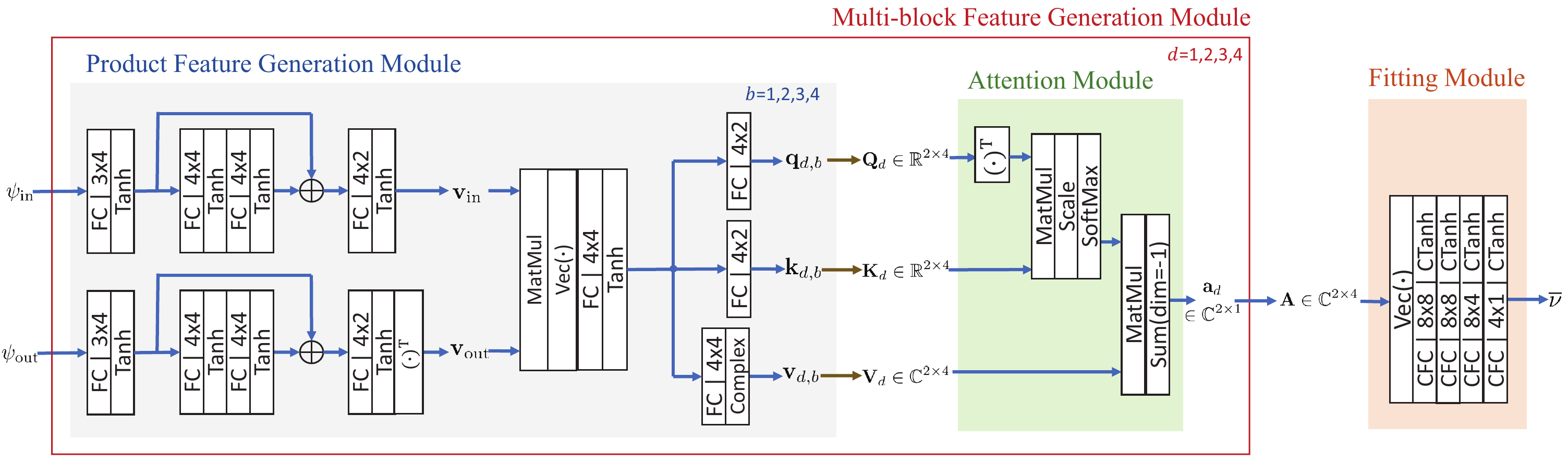}
\caption{\color{black} Architecture of the proposed DNN. Note that ``FC 4$\times$2'' means a fully-connected layer with 4 input channel and 2 output channel, ``CFC 4$\times$2'' means a complex fully-connected layer,, ``CTanh'' means a complex activation layer using Tanh($\cdot$) function, ``MatMul'' means matrix multiplication, ``$(\cdot)^{\rm T}$'' means transpose a matrix, and ``Vec($\cdot$)'' means vectorization.}
\label{dnn_major}
\end{figure*}

The overall DNN design is illustrated in {\figurename~\ref{dnn_major}}. 
To reduce complexity, the proposed DNN is designed as a two-tier parallel-block structure.

\begin{table*}[!t]
\footnotesize
\renewcommand{\arraystretch}{1.1}
\caption{NMSE of the RIS Response Function Approximations.}
\label{table_nmse}
\centering
\begin{tabular}{|c|c|c|c|c|c|c|c|c|}
\hline
     & State 1 & State 2 & State 3 & State 4 & State 5 & State 6 & State 7 & State 8
     \\ \hline
Kernel & 5.98e-3  & 3.92e-3 & 2.33e-3 & 4.63e-3 & 2.62e-3 & 3.86e-3 & 3.25e-3 & 3.12e-3
     \\ \hline
DNN  & 2.46e-5 & 5.91e-5 & 2.1e-6 & 5.8e-6 & 1.33e-5 & 1.48e-5 & 6.0e-6 & 2.1e-6
     \\ \hline
\hline
    & State 9 & State 10 & State 11 & State 12 & State 13 & State 14 & State 15 & State 16
     \\ \hline
Kernel & 3.80e-3 & 6.15e-3 & 3.16e-3 & 1.36e-3 & 5.13e-3 & 1.22e-2 & 3.43e-3 & 1.45e-2
     \\ \hline
DNN  & 3.3e-6 & 6.3e-6 & 2.04e-5 & 1.22e-5 & 2.3e-6 & 5.3e-6 & 3.4e-6 & 1.13e-5
    \\ \hline
\end{tabular}
\end{table*}

The inner tier includes a ``Product Feature Generation Module'' for parallel feature generation and an ``Attention Module'' to reduce the dimensionality of the stacked features.
\begin{itemize}
\item {\color{black}{\bf Product Feature Generation Module}: This module comprises four parallel blocks ($b=1,2,3,4$) with identical DNN structures but different weights. In each block, the output feature is generated using a product structure for the two data links, with inputs input $\psi_{\rm in}$ and $\psi_{\rm out}$ processed separately. The DNN function is defined by
    \begin{equation}\label{equ:product_feature_module}
    \begin{aligned}[b]
    {\bf q}_{d,b}, {\bf k}_{d,b}, {\bf v}_{d,b} = f_{{\rm PF},d,b}(\psi_{\rm in},\psi_{\rm out}; {\bm \omega}_{{\rm PF},d,b}),
    \end{aligned}
    \end{equation}
    where ${\bm \omega}_{{\rm PF},b}$ is the DNN weights, and the overall weights is ${\bm \omega}_{{\rm PF},d}=\{{\bm \omega}_{{\rm PF},d,b}, \forall b\}$.}
\item {\bf Attention Module}: This module reduces the dimension of the stacked feature ${\bf V}_{d}$ to a vector ${\bf a}_{d}$:
    \begin{equation}\label{equ:attention_module}
    \begin{aligned}[b]
    {\bf a}_{d} = f_{{\rm AT},d}({\bf Q}_{d}, {\bf K}_{d}, {\bf V}_{d}; {\bm \omega}_{{\rm AT},d}),
    \end{aligned}
    \end{equation}
    where ${\bm \omega}_{{\rm AT},d}$ denotes the DNN weights.
\end{itemize}

The outer tier includes a ``Multi-block Feature Generation Module'' for parallel feature generation and a ``Fitting Module'' to reduce the dimensionality of the stacked features.
\begin{itemize}
\item {\bf Multi-block Feature Generation Module}: This module comprises four parallel blocks ($b=1,2,3,4$) with identical DNN structures but different weights. Each block is designed using to ``inner tier'' structure introduced above:
    \begin{equation}\label{equ:multi_feature_module}
    \begin{aligned}[b]
    {\bf a}_{d} = f_{{\rm FG},d}(\psi_{\rm in},\psi_{\rm out}; {\bm \omega}_{{\rm FG},d}),
    \end{aligned}
    \end{equation}
    where the weights ${\bm \omega}_{{\rm FG},d}=\{{\bm \omega}_{{\rm PF},d}, {\bm \omega}_{{\rm AT},d}\}$, and the overall weights are denoted by ${\bm \omega}_{{\rm FG}}=\{{\bm \omega}_{{\rm FG},d}, \forall d\}$.
\item {\bf Fitting Module}: This module is simply designed as the cascaded of complex fully-connected layers to get the finally output:
    \begin{equation}\label{equ:fitting_module}
    \begin{aligned}[b]
    \overline{\nu} = f_{{\rm FG}}({\bf A}; {\bm \omega}_{{\rm FM}}),
    \end{aligned}
    \end{equation}
    where ${\bm \omega}_{{\rm FM}}$ summarizes the DNN weights. The overall DNN weights are ${\bm \omega}_\ell=\{{\bm \omega}_{{\rm FG}}, {\bm \omega}_{{\rm FM}} \}$
\end{itemize}

\underline{Discussion}: {\color{black}Both the ``Attention Module'' and the ``Fitting Module'' can reduce feature dimensions; however, the ``Attention Module'' typically offers better performance but requires additional feature-related inputs. In contrast, the ``Fitting Module'' has a simpler structure, though its performance may be slightly lower. The proposed two-tier parallel-block structure effectively balances complexity and performance by applying these two types of modules in separate tiers.}

\subsection{Training and Results}
For both the product-Legendre-kernel-based solution and the DNN-based solution, we train the weights ${\bm \omega}_\ell$ using the same loss function to minimize the approximation NMSE:
\begin{align}
{\cal P}{\text{(B)}} \; \min_{{\bm \omega}_\ell}& \; \frac{1}{J} \sum_{j=1}^J \frac{| \overline{\nu}(\phi_{\rm in}^{(j)},\phi_{\rm out}^{(j)}; {\bm \omega}_\ell)
-\nu^{(j)}|^2}
{|\nu^{(j)}|^2},
\end{align}
where $J$ is the size of the training dataset.  The training dataset is generated by randomly selecting $10^7$ pairs of incident and reflection angles for each RIS state. Likewise, the validation and testing sets consist of $10^6$ samples each. ADAM optimizer is employed to solve problem ${\cal P}{\text{(B)}}$ with a learning rate of $10^{-4}$. The batch size is defined as $2048$, and the maximum number of training epochs is set to be $1000$.

\begin{table}[!ht]
\footnotesize
\renewcommand{\arraystretch}{1.1}
\caption{Complexity of the Proposed Two Solutions.}
\label{complexity}
\centering
\begin{tabular}{|c|c|c|}
\hline
         &  Kernel (B=5) &  DNN   \\ \hline
FLOPs &  4530       & 1432          \\ \hline
Parameters &  625       & 550        \\ \hline
Memory (KB) &  4.6       & 196      \\ \hline
\end{tabular}
\end{table}

We show the complexity of our two proposed solutions in Table \ref{complexity}, detailing the number of DNN parameters, FLOPs, and memory needed for storage. Both solutions exhibit low computational and memory complexity. Specifically, the computational complexity of the DNN is significantly lower, while the Kernel-based solution requires less memory for storage.

We present the approximate NMSE for our proposed solutions in Table \ref{table_nmse}. The product-Legendre-kernel-based solution achieves an NMSE of around 0.001 in most cases but performs poorly for states 14 and 16. In contrast, the DNN solution consistently performs well, reducing the NMSE by about 100 times compared to the product-Legendre-kernel-based solution.



\section{Model-based Channel Estimation for Pixel-based-RIS-Assisted MU-MISO system}
\subsection{Transmission Frames for RIS Sounding and Data Transmission}
\begin{figure}
[!ht]
\centering
\includegraphics[width=.85\columnwidth]{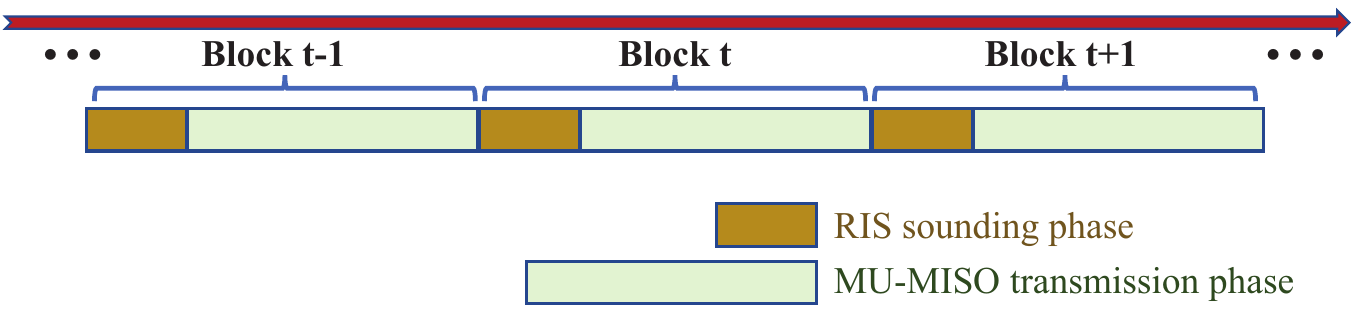}
\caption{Illustration of the transmission blocks in a timeline.}
\label{frame_timeline}
\end{figure}

We utilize a transmission frame structure that separately supports RIS channel sounding and MU-MISO transmission, as illustrated in  {\figurename~\ref{frame_timeline}}. Specifically, each time block is divided into two transmission phases.

\begin{figure}
[!ht]
\centering
\includegraphics[width=.85\columnwidth]{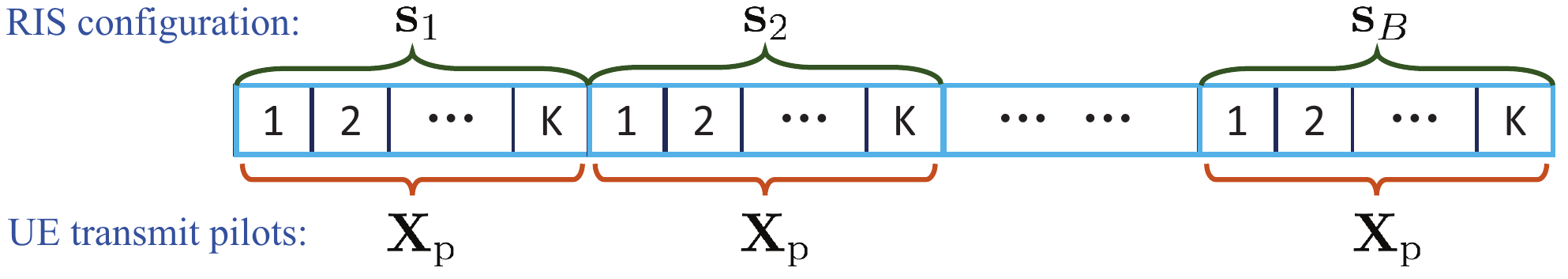}
\caption{Illustration of the transmission in the sounding phase.}
\label{frame_sounding}
\end{figure}

\subsubsection{RIS Sounding Phase}
In this phase, we use a similar uplink channel sounding protocol for traditional RIS cascaded channel estimation \cite{RISCE2020twc,RISCE2022Guo,RISCE2022twc,RISCE2020JSAC,RISCE2024TWC,RISCE2024TCOM,Chenjie2023RISCE}, as shown in {\figurename~\ref{frame_sounding}}. The BS receives $KB$ uplink sounding pilots from users, and the RIS elements are configured into $B$ different states based on ${\bf s}_1, {\bf s}_2, \cdots, {\bf s}_B$. The users repeatedly transmit an orthogonal pilot matrix ${\bf X}_{\rm p}=[{\bf x}_{{\rm p},1},{\bf x}_{{\rm p},2},\cdots,{\bf x}_{{\rm p},K}]$ a total of $B$ times. The received signal is given by
\begin{equation}\label{equ:signal_sounding_phase}
\begin{aligned}[b]
\dot{\bf Y}_{{\rm u},t,b}&=\sum_{k=1}^K {\bf h}_{k,t}({\bf s}_b) {\bf x}_{{\rm p},k}^{\rm T}+ \dot{\bf Z}_{{\rm u},t,b}\\
&={\bf H}_{t}({\bf s}_b) {\bf X}_{{\rm p}}^{\rm T}+ \dot{\bf Z}_{{\rm u},t,b},
\end{aligned}
\end{equation}
where ${\bf H}_{t}({\bf s}_b)=[{\bf h}_{1,t}({\bf s}_b),{\bf h}_{2,t}({\bf s}_b),\cdots,{\bf h}_{K,t}({\bf s}_b)]$ for $b=1,2,\cdots,B$, and $\dot{\bf Z}_{{\rm u},t,b}$ represents the received AWGN with elements following i.i.d. ${\cal{CN}}(0,\sigma_{\rm u}^2)$. Since ${\bf X}_{\rm p}$ is an orthogonal matrix, we can remove it from \eqref{equ:signal_sounding_phase}, yielding:
\begin{equation}\label{equ:signal_sounding_phase_2}
\begin{aligned}[b]
{\bf Y}_{{\rm u},t,b}&
&={\bf H}_{t}({\bf s}_b)+ {\bf Z}_{{\rm u},t,b},
\end{aligned}
\end{equation}
where ${\bf Y}_{{\rm u},t,b} = {\bf x}_{{\rm p},k}^\star \dot{\bf Y}_{{\rm u},t,b}$ and ${\bf Z}_{{\rm u},t,b} = {\bf x}_{{\rm p},k}^\star \dot{\bf Z}_{{\rm u},t,b}$.

\subsubsection{MU-MISO Transmission Phase}
In this phase, the RIS is configured to state ${\bf{s}}_t^\ast$ by solving  ${\mathcal{P}}{(\text{A})}$ mentioned in Section \ref{sec:model_MU_MISO}. The downlink received signal at user $k$ becomes:
\begin{equation}\label{equ:signal_user_k_star}
\begin{aligned}[b]
y_{k,t}&={\bf h}_{k,t}({\bf s}_t^\ast)^{\rm T} \sum_{k=1}^{K} {\bf w}_{k,t} x_{k,t}+ z_{k,t},
\end{aligned}
\end{equation}
where the precoder ${\bf w}_{k,t}$ has been defined in \eqref{equ:precoder_rzf}.

\subsection{Estimating and Predicting Instantaneous Channels with an Approximate Model for Dominant Rays}
Recall the cascaded channel model in \eqref{equ:channel_cascaded}, reproduced here:
\begin{equation}\label{equ:channel_cascaded_copy}
\begin{aligned}[b]
{\bf h}_{k,t}({\bf s})
=&\sum_{i,j} {\bf G}_{t}^{(i)} {\rm diag} \left({\bm \nu}(\phi_{\rm in}^{(i)},\phi_{{\rm out},k}^{(j)};{\bf s})\right) {\bf h}_{{\rm r},k,t}^{(j)}.
\end{aligned}
\end{equation}
Exactly predicting ${\bf h}_{k,t}({\bf s})$ for any ${\bf s}$ is difficult due to the non-separable state response and the numerous rays in wireless propagation. However, passive beamforming design can tolerate some prediction errors, allowing for an approximate model that enables tractable channel estimation.

\subsubsection{Approximate Model for BS-RIS Channel}
The BS-RIS channel model in Section \ref{sec:system_model_scattering} can be simplified for two reasons. {\color{black}First, the BS uses a conventional antenna array, and its response is independent of the RIS configuration. Second, the wireless propagation rays are clustered around the dominant scatterers in the environment. Therefore, we can represent the rays associated with the same scatterer using one or two dominant rays with a small approximation loss. The proposed approximate model is given by
\begin{equation}\label{equ:new_model_G}
\begin{aligned}[b]
{\bf G}_{t} &= \sum_{m=1}^M \sum_{i=0}^I a_{t,m,i} {\bf v}_m {\bm \alpha}^{\rm H}_{\rm RIS}({\overline \phi}_{\rm in}^{(i)}),
\end{aligned}
\end{equation}
where ${\bf v}_m$ is the $m$-th column of an orthogonal matrix $\bf V$, defined as
\begin{equation}\label{equ:bf_V}
\begin{aligned}[b]
{\bf V}&={\rm{diag}}({\bm \alpha}_{\rm BS}(\phi_{\rm BS}^{(0)})) \left( {\rm{DFT}}(M_1) \otimes {\rm{DFT}}(M_2) \right),
\end{aligned}
\end{equation}
for a $M_1 \times M_2$ UPA, with ${\bf v}_1={\bm \alpha}_{\rm BS}(\phi_{\rm BS}^{(0)})$.} Without loss of generality, we set $a_{t,1,0}=1$ for all $t$. In addition, since the LoS BS-RIS link is know, we have ${\overline \phi}_{\rm in}^{(0)}={\phi}_{\rm in}^{(0)}$, and we assume ${\overline \phi}_{\rm in}^{(i)} \in {\cal D}$ for all $i$, where $\cal D$ is a grid-based angle set.

\subsubsection{Approximate Model for RIS-UE Channels}
Similarly, the RIS-UE channels can be approximated by dominated rays with angles on the grid $\cal D$:
\begin{equation}\label{equ:new_model_Hr}
\begin{aligned}[b]
{\bf h}_{{\rm r},k,t} & = \sum_{j=0}^J b_{k,t,j} {\bm \alpha}_{\rm RIS}({\overline \phi}_{{\rm out},k}^{(j)}),
\end{aligned}
\end{equation}
where ${\overline \phi}_{{\rm out},k}^{(j)} \in {\cal D}$ for all $j$ and $k$.

\subsubsection{Approximate Cascaded Channels}
Combining \eqref{equ:new_model_G} and \eqref{equ:new_model_Hr}, the approximate cascaded channel is expressed as
\begin{equation}\label{equ:channel_cascaded_app}
\begin{aligned}[b]
{\bf h}_{k,t}({\bf s})
=&\sum_{m=1}^M \sum_{i=0}^I \sum_{j=0}^J a_{t,m,i} d_{k,t,j} {\bf v}_m {\bm \alpha}^{\rm H}_{\rm RIS}({\overline \phi}_{\rm in}^{(i)}) \\
&\quad {\rm diag} \left({\bm \nu}({\overline \phi}_{\rm in}^{(i)},{\overline \phi}_{{\rm out},k}^{(j)};{\bf s})\right)
{\bm \alpha}_{\rm RIS}({\overline \phi}_{{\rm out},k}^{(j)}).
\end{aligned}
\end{equation}
To reconstruct ${\bf h}_{k,t}({\bf s})$, we need to estimate the angles ${\overline \phi}_{\rm in}^{(i)}$ and ${\overline \phi}_{{\rm out},k}^{(j)}$ for the dominant rays, along with the time-varying channel responses $a_{t,m,i}$ and  $d_{k,t,j}$.

\subsection{\color{black}Instantaneous Channel Estimation Given Knowledge of Angles}\label{sec:CE_inst}
The cascaded channel in \eqref{equ:channel_cascaded_app} is further written by
\begin{equation}\label{equ:channel_cascaded_app_2}
\begin{aligned}[b]
{\bf h}_{k,t}({\bf s}_b)
=&\sum_{m=1}^M \sum_{i=0}^I \sum_{j=0}^J a_{t,m,i} d_{k,t,j} {f}_{b,i,j,k} {\bf v}_m ,
\end{aligned}
\end{equation}
where the parameters ${f}_{b,m,i,j}$ is defined as
\begin{align}
{f}_{b,i,j,k}
=&   {\bm \alpha}^{\rm H}_{\rm RIS}({\overline \phi}_{\rm in}^{(i)}) \notag \\
&\quad {\rm diag} \left({\bm \nu}({\overline \phi}_{\rm in}^{(i)},{\overline \phi}_{{\rm out},k}^{(j)};{\bf s}_b)\right)
{\bm \alpha}_{\rm RIS}({\overline \phi}_{{\rm out},k}^{(j)}), \label{equ:channel_cascaded_app_basis}
\end{align}
We present a three-step method to estimate the instantaneous channel responses $a_{t,m,i}$ and  $d_{k,t,j}$ given knowledge of all ${f}_{b,i,j,k}$.

\begin{itemize}
\item {{\bf Step 1} (Estimating LoS Link)}: Since the LoS ray is generally stronger than each NLoS ray, we first estimate $d_{k,t,0}$ for all $k$ given $a_{t,1,0}=1$ from observations ${\bf Y}_{{\rm u},t,b}$ in \eqref{equ:signal_sounding_phase_2} by treating all the NLoS components as interference:
    \begin{equation}\label{equ:obs_los_est}
    \begin{aligned}[b]
    {\bf y}_{{\rm u},t,b,k}
    =& d_{k,t,0}  {f}_{b,0,0,k} {\bf v}_{0}+{\bf e}_{t,b,k}+{\bf z}_{{\rm u},t,b,k},
    \end{aligned}
    \end{equation}
    where ${\bf e}_{t,b,k}$ summarizes all the NLoS components. Then, $d_{k,t,0}$ is estimated by the least squares (LS) method:
    \begin{equation}\label{equ:obs_los_est_LS}
    \begin{aligned}[b]
    d_{k,t,0}
    =& \frac{ \sum_b {f}_{b,0,0,k}^\star {\bf v}_{0}^{\rm H} {\bf y}_{{\rm u},t,b,k} } { \sum_b |{f}_{b,0,0,k}|^2  }.
    \end{aligned}
    \end{equation}
\item {{\bf Step 2} (Estimating All $a_{t,m,i}$)}: Since $\bf V$ is orthogonal matrix, we have the observation for the $m$-th angular direction at the BS:
    \begin{equation}\label{equ:obs_los_cancel}
    \begin{aligned}[b]
    \overline{y}_{{\rm u},t,b,k,m}
    =& {\bf v}_{m}^{\rm H} {\bf y}_{{\rm u},t,b,k}  \\
    =&  \sum_{i=0}^I  a_{t,m,i} {f}_{b,i,0,k}  +\overline{\bf e}_{t,b,k}+\overline{\bf z}_{{\rm u},m,t,b,k},
    \end{aligned}
    \end{equation}
    where $\overline{\bf e}_{t,b,k}$ summarizes all the components containing the RIS-UE NLoS links, and $\overline{\bf z}_{{\rm u},m,t,b,k}={\bf v}_{m}^{\rm H} {\bf z}_{{\rm u},t,b,k}$. Letting $\overline{\bf y}_{{\rm u},t,m}=[\overline{y}_{{\rm u},t,1,1,m},\overline{y}_{{\rm u},t,1,2,m},\cdots,\overline{y}_{{\rm u},t,B,K,m}]$, ${\bf f}_{i,0}=[{f}_{1,i,0,1},{f}_{1,i,0,2},\cdots,{f}_{B,i,0,K}]$, and ${\bf a}_{t,m}=[a_{t,m,0},a_{t,m,1},\cdots,a_{t,m,I}]$, we have estimation:
    \begin{equation}\label{equ:obs_bs_ris_nlos_est_LS}
    \begin{aligned}[b]
    {\bf a}_{t,m}
    =& \left( \overline{\bf F}_{0}^{\rm H} \overline{\bf F}_{0} \right)^{-1} \overline{\bf F}_{0}^{\rm H} \overline{\bf y}_{{\rm u},t,m} ,
    \end{aligned}
    \end{equation}
    where $\overline{\bf F}_{0}=[{\bf f}_{1,0},{\bf f}_{2,0},\cdots,{\bf f}_{I,0}]$.
\item {{\bf Step 3} (Estimating All $d_{k,t,j}$)}: Since all BS-RIS parameters $a_{t,m,i}$ have been estimated, we define
    \begin{equation}\label{equ:obs_irs_use_basis}
    \begin{aligned}[b]
    \widetilde{\bf v}_{k,t,j,b}
    =& \sum_{m=1}^M \sum_{i=0}^I  a_{t,m,i} {f}_{b,i,j,k} {\bf v}_m ,
    \end{aligned}
    \end{equation}
    and the received pilot observation is written by
    \begin{equation}\label{equ:obs_los_est_new3}
    \begin{aligned}[b]
    {\bf y}_{{\rm u},t,b,k}
    =& \sum_{j=0}^J \widetilde{\bf v}_{k,t,j,b} d_{k,t,j}  +{\bf z}_{{\rm u},t,b,k}\\
    =& \widetilde{\bf V}_{k,t,b} {\bf d}_{k,t}  +{\bf z}_{{\rm u},t,b,k},
    \end{aligned}
    \end{equation}
    where ${\bf d}_{k,t}=[d_{k,t,0},d_{k,t,1},\cdots,d_{k,t,J}]^{\rm T}$ and $\widetilde{\bf V}_{k,t,b}=[\widetilde{\bf v}_{k,t,0,b},\widetilde{\bf v}_{k,t,1,b},\cdots,\widetilde{\bf v}_{k,t,J,b}]$. Then, we finally have estimation:
    \begin{equation}\label{equ:obs_ris_ue_nlos_est_LS}
    \begin{aligned}[b]
    {\bf d}_{k,t}
    =& \left( \sum_{b=1}^B \widetilde{\bf V}_{k,t,b}^{\rm H} \widetilde{\bf V}_{k,t,b} \right)^{-1}
     \left(\sum_{b=1}^B \widetilde{\bf V}_{k,t,b}^{\rm H} {\bf y}_{{\rm u},t,b,k} \right).
    \end{aligned}
    \end{equation}
\end{itemize}

{\color{black}We summarize the estimation method in Algorithm \ref{alg:Instan_CE}. The complexity of Algorithm \ref{alg:Instan_CE} is ${\cal O}(M I^3+M I^2 B K+K J^3+M B K J^2 )$, which is independent of the angle set size $|{\cal D}|$.} Additionally, to ensure the feasibility of the matrix inverse operations in \eqref{equ:obs_bs_ris_nlos_est_LS} and \eqref{equ:obs_ris_ue_nlos_est_LS}, we require $I < BK$ and $J < BM$  for the proposed approximate model in \eqref{equ:channel_cascaded_app}.

\begin{algorithm}[!t]
\caption{Proposed Three-Step Estimation on the Instantaneous Channel }\label{alg:Instan_CE}
\begin{algorithmic}[1]
\State\textbf{Input}:  Observation set ${\bf Y}_{{\rm u},t,b}$ for all $b$ and the angles;
\State  Estimate $d_{k,t,0}$ by \eqref{equ:obs_los_est_LS} for LoS;
\State  Estimate ${\bf a}_{t,m}$ for all $m$ by \eqref{equ:obs_bs_ris_nlos_est_LS};
\State  Estimate ${\bf d}_{k,t}$ for all $k$ by \eqref{equ:obs_ris_ue_nlos_est_LS};
\State\textbf{Output}: Predicting  ${\bf h}_{k,t}({\bf s})$ based on \eqref{equ:channel_cascaded_app_2}.
\end{algorithmic}
\end{algorithm}

\subsection{Statistical Knowledge Estimation over $T$ Historical Observation Blocks}
In this subsection, we focus on estimating the angles ${\overline \phi}_{\rm in}^{(i)}$ and ${\overline \phi}_{{\rm out},k}^{(j)}$ for the dominant rays using historical observations ${\cal Y}_t=\{ {\bf Y}_{{\rm u},t,b}, \; t=1,2,\cdots,T \}$. Similar to instantaneous channel estimation, we accomplish this through a three-step method:
\begin{itemize}
\item {{\bf Step 1} (Estimating LoS Angles ${\overline \phi}_{{\rm out},k}^{(0)}$)}: Since ${\overline \phi}_{\rm in}^{(0)}$ is known, we can search the strongest angle in the grid set $\cal D$ as the estimation of ${\overline \phi}_{{\rm out},k}^{(0)}$. Specifically, assuming ${\overline \phi}_{{\rm out},k}^{(0)}={\phi}$, the LoS link is given by
    \begin{equation}\label{equ:stat_channel_los}
    \begin{aligned}[b]
    {\bf h}_{k,t}^{(0)}({\bf s}_b,{\phi})
    =& d_{k,t,0}({\phi}) {\bf v}_0 {\bm \alpha}^{\rm H}_{\rm RIS}({\overline \phi}_{\rm in}^{(0)}) \\
    &\quad {\rm diag} \left({\bm \nu}({\overline \phi}_{\rm in}^{(0)},{\phi} ;{\bf s}_b)\right)
    {\bm \alpha}_{\rm RIS}({\phi}),
    \end{aligned}
    \end{equation}
    and $d_{k,t,0}({\phi})$ is defined based on \eqref{equ:obs_los_est_LS}:
    \begin{equation}\label{equ:stat_obs_los_est_LS}
    \begin{aligned}[b]
    d_{k,t,0}({\phi})
    =& \frac{ \sum_b  {\bf h}_{k,t}^{(0)}({\bf s}_b,{\phi})^{\rm H} {\bf y}_{{\rm u},t,b,k} }
    { \sum_b {\bf h}_{k,t}^{(0)}({\bf s}_b,{\phi})^{\rm H} {\bf h}_{k,t}^{(0)}({\bf s}_b,{\phi})  }.
    \end{aligned}
    \end{equation}
    Then, ${\overline \phi}_{{\rm out},k}^{(0)}$ is estimated by solving:
    \begin{subequations}
    \begin{align}
        {\mathcal{P}}{(\text{C-1})}
      \; {\overline \phi}_{{\rm out},k}^{(0)}  & =\; \underset{{\phi} \in {\cal D} }{\rm{argmin}}   \;
    \sum_{t,b} \| {\bf y}_{{\rm u},t,b,k}-{\bf h}_{k,t}^{(0)}({\bf s}_b,{\phi}) \|_2^2 . \notag
    \end{align}
    \end{subequations}
\item {{\bf Step 2} (Estimating BS-RIS NLoS Angles ${\overline \phi}_{\rm in}^{(i)}$)}: We initialize the BS-RIS angle set ${\cal A}_{\rm in}=\{{\overline \phi}_{\rm in}^{(0)}\}$. Given the RIS-UE LoS angles ${\overline \phi}_{{\rm out},k}^{(0)}$ and the response $d_{k,t,0}$, we assume the incident angle of the largest NLoS ray is ${\overline \phi}_{\rm in}^{(i)}=\phi$, and the cascaded channel for this ray pair can be expressed as follows
    \begin{equation}\label{equ:stat_channel_cascaded_app_2}
    \begin{aligned}[b]
    {\bf h}_{k,t}^{(1)}({\bf s}_b,\phi)
    =&\sum_{m=1}^M  a_{t,m}(\phi) d_{k,t,0} {\bf v}_m {\bm \alpha}^{\rm H}_{\rm RIS}(\phi) \\
    &\quad {\rm diag} \left({\bm \nu}(\phi,{\overline \phi}_{{\rm out},k}^{(0)};{\bf s}_b)\right)
    {\bm \alpha}_{\rm RIS}({\overline \phi}_{{\rm out},k}^{(0)})\\
    =& \sum_{m=1}^M  a_{t,m}(\phi) {\bf v}_m \overline{f}_{b,k}(\phi),
    \end{aligned}
    \end{equation}
    where $\overline{f}_{b,k}(\phi)$ is defined as follows
    \begin{align}
    \overline{f}_{b,k,t}(\phi)
    =&   d_{k,t,0} {\bm \alpha}^{\rm H}_{\rm RIS}(\phi)\notag \\
    &\; {\rm diag} \left({\bm \nu}(\phi,{\overline \phi}_{{\rm out},k}^{(0)};{\bf s}_b)\right)
    {\bm \alpha}_{\rm RIS}({\overline \phi}_{{\rm out},k}^{(0)}). \label{equ:stat_channel_cascaded_app_basis}
    \end{align}
    Then, based on the estimator in \eqref{equ:obs_los_est_LS}, \eqref{equ:obs_los_cancel} and \eqref{equ:obs_bs_ris_nlos_est_LS}, $a_{t,m}(\phi)$ in \eqref{equ:stat_channel_cascaded_app_2} can be estimated by
    \begin{equation}\label{equ:stat_obs_bs_ris_nlos_est_LS}
    \begin{aligned}[b]
    a_{t,m}(\phi)
    =&  \frac{ \sum_{b,k} \overline{f}_{b,k,t}(\phi)^\star {\bf v}_{m}^{\rm H} \widetilde{\bf y}_{{\rm u},t,b,k} }
    { \sum_{b,k} | \overline{f}_{b,k,t} |^2 },
    \end{aligned}
    \end{equation}
    where $\widetilde{\bf y}_{{\rm u},t,b,k}$ is initialized by
    \begin{equation}\label{equ:init_y}
    \begin{aligned}[b]
    \widetilde{\bf y}_{{\rm u},t,b,k}
    =&{\bf y}_{{\rm u},t,b,k}-{\bf h}_{k,t}^{(1)}({\bf s}_b,{\overline \phi}_{\rm in}^{(0)}).
    \end{aligned}
    \end{equation}
    The strongest incident angle ${\overline \phi}_{\rm in}^{(i)}$ for $\widetilde{\bf y}_{{\rm u},t,b,k}$ is estimated by solving:
    \begin{subequations}
    \begin{align}
        {\mathcal{P}}{(\text{C-2})}
      \; {\overline \phi}_{\rm in}^{(i)}  & =\; \underset{{\phi} \in {\cal D} }{\rm{argmin}}   \;
    \sum_{t,b,k} \| \widetilde{\bf y}_{{\rm u},t,b,k}-{\bf h}_{k,t}^{(1)}({\bf s}_b,{\phi}) \|_2^2 , \notag
    \end{align}
    \end{subequations}
    and we update ${\cal A}_{\rm in}={\cal A}_{\rm in} \cup \{{\overline \phi}_{\rm in}^{(i)}\}$.
    We next update the observation by
    \begin{equation}\label{equ:stat_obs_bs_ris_nlos_ite}
    \begin{aligned}[b]
    \widetilde{\bf y}_{{\rm u},t,b,k}
    =&  \widetilde{\bf y}_{{\rm u},t,b,k}-\widetilde{\bf h}_{k,t}^{(1)}({\bf s}_b; {\cal A}_{\rm in}),
    \end{aligned}
    \end{equation}
    to search for the strongest incident angle that is not in ${\cal A}_{\rm in}$,
    where $\widetilde{\bf h}_{k,t}^{(1)}({\bf s}_b; {\cal A}_{\rm in})$ is defined by
    \begin{equation}\label{equ:channel_cascaded_app_temp1}
    \begin{aligned}[b]
    \widetilde{\bf h}_{k,t}^{(1)}({\bf s}_b; {\cal A}_{\rm in})
    =&\sum_{m=1}^M \sum_{i=0}^{|{\cal A}_{\rm in}|-1} a_{t,m,i} d_{k,t,0} {f}_{b,i,0,k} {\bf v}_m,
    \end{aligned}
    \end{equation}
    whose parameters $\{a_{t,m,i}\}$ are calculated by \eqref{equ:obs_bs_ris_nlos_est_LS}.

\begin{algorithm}[!t]
\caption{Algorithm for Channel Statistical Knowledge Estimation}\label{alg:CE_Stat}
\begin{algorithmic}[1]
\State\textbf{Input}:  Observation  ${\cal Y}_t$;
\State Estimate LoS RIS-UE angles ${\overline \phi}_{{\rm out},k}^{(0)}$ by solving ${\mathcal{P}}{(\text{C-1})}$;
\State Initialize $\widetilde{\bf y}_{{\rm u},t,b,k}$ by \eqref{equ:init_y} and ${\cal A}_{\rm in}=\{{\overline \phi}_{\rm in}^{(0)}\}$;
\While{${\mathbb E}_{t,b,k}\| \widetilde{\bf y}_{{\rm u},t,b,k} \|_2^2 > M \sigma_{\rm u}^2 $ and $|{\cal A}_{\rm in}|<BK$}
\State Search ${\overline \phi}_{\rm in}^{(i)}$ by solving ${\mathcal{P}}{(\text{C-2})}$;
\State Update ${\cal A}_{\rm in}={\cal A}_{\rm in} \cup \{{\overline \phi}_{\rm in}^{(i)}\}$;
\State Update $\widetilde{\bf y}_{{\rm u},t,b,k}$ \eqref{equ:stat_obs_bs_ris_nlos_ite};
\EndWhile
\State Initialize $\widetilde{\bf y}_{{\rm u},t,b,k}={\bf y}_{{\rm u},t,b,k}$ and ${\cal A}_{{\rm out},k}= \emptyset$;
\For{$k=1:K$}
\While{${\mathbb E}_{t,b,k}\| \widetilde{\bf y}_{{\rm u},t,b} \|_2^2 > M \sigma_{\rm u}^2 $ and $|{\cal A}_{{\rm out},k}|<BM$}
\State Search ${\overline \phi}_{{\rm out},k}^{(j)}$ by solving ${\mathcal{P}}{(\text{C-3})}$;
\State Update ${\cal A}_{{\rm out},k}={\cal A}_{\rm in} \cup \{{\overline \phi}_{{\rm out},k}^{(j)}\}$;
\State Update $\widetilde{\bf y}_{{\rm u},t,b,k}$ \eqref{equ:stat2_obs_bs_ris_nlos_ite};
\EndWhile
\EndFor
\State\textbf{Output}: The estimated angles sets ${\cal A}_{\rm in}$ and ${\cal A}_{{\rm out},k}$ for the dominant rays.
\end{algorithmic}
\end{algorithm}

\item {{\bf Step 3} (Estimating RIS-UE Angles ${\overline \phi}_{{\rm out},k}^{(j)}$)}: We estimate the reflection angle ${\overline \phi}_{{\rm out},k}^{(j)}$ in this step given ${\cal A}_{\rm in}$ and response parameters $\{a_{t,m,i}\}$. We initialize the RIS-UE angle set by ${\cal A}_{{\rm out},k}= \emptyset$, and assume the strongest angle is ${\overline \phi}_{{\rm out},k}^{(j)}=\phi$. Then, the cascaded channel related to this angle is given by
    \begin{equation}\label{equ:stat2_channel_cascaded_app}
    \begin{aligned}[b]
    {\bf h}_{k,t}^{(2)}({\bf s}_b,\phi)
    =&\sum_{m=1}^M \sum_{i=0}^I  a_{t,m,i} d_{k,t}(\phi) {\bf v}_m {\bm \alpha}^{\rm H}_{\rm RIS}({\overline \phi}_{\rm in}^{(i)}) \\
    &\quad {\rm diag} \left({\bm \nu}({\overline \phi}_{\rm in}^{(i)},\phi;{\bf s}_b)\right)
    {\bm \alpha}_{\rm RIS}(\phi)\\
    =& d_{k,t}(\phi) \widetilde{\bf f}_{t,b},
    \end{aligned}
    \end{equation}
    where $\widetilde{\bf f}_{t,b}$ is defined by
    \begin{equation}
    \begin{aligned}[b]
    \widetilde{\bf f}_{t,b}
    =&   \sum_{m=1}^M \sum_{i=0}^I  a_{t,m,i} {\bf v}_m {\bm \alpha}^{\rm H}_{\rm RIS}({\overline \phi}_{\rm in}^{(i)}) \\
    &\quad {\rm diag} \left({\bm \nu}({\overline \phi}_{\rm in}^{(i)},\phi;{\bf s}_b)\right)
    {\bm \alpha}_{\rm RIS}(\phi),
    \end{aligned}
    \end{equation}
    and $d_{k,t}(\phi)$ is given by
    \begin{equation}\label{equ:stat2_obs_bs_ris_nlos_est_LS}
    \begin{aligned}[b]
    d_{k,t}(\phi)
    =&  \frac{ \sum_{b} \widetilde{\bf f}_{t,b}^{\rm H} \widetilde{\bf y}_{{\rm u},t,b,k} }
    { \sum_{b} \widetilde{\bf f}_{t,b}^{\rm H} \widetilde{\bf f}_{t,b} },
    \end{aligned}
    \end{equation}
    with $\widetilde{\bf y}_{{\rm u},t,b,k}$ initialized by ${\bf y}_{{\rm u},t,b,k}$.
    Then, we solve ${\overline \phi}_{{\rm out},k}^{(j)}$ by:
    \begin{subequations}
    \begin{align}
        {\mathcal{P}}{(\text{C-3})}
      \; {\overline \phi}_{{\rm out},k}^{(j)}  & =\; \underset{{\phi} \in {\cal D} }{\rm{argmin}}   \;
    \sum_{t,b} \| \widetilde{\bf y}_{{\rm u},t,b,k}-{\bf h}_{k,t}^{(2)}({\bf s}_b,{\phi}) \|_2^2 , \notag
    \end{align}
    \end{subequations}
    and we update ${\cal A}_{{\rm out},k}={\cal A}_{{\rm out},k} \cup \{{\overline \phi}_{{\rm out},k}^{(j)}\}$.
    We next update the observation by
    \begin{equation}\label{equ:stat2_obs_bs_ris_nlos_ite}
    \begin{aligned}[b]
    \widetilde{\bf y}_{{\rm u},t,b,k}
    =&  \widetilde{\bf y}_{{\rm u},t,b,k}-\widetilde{\bf h}_{k,t}^{(2)}({\bf s}_b; {\cal A}_{{\rm out},k}),
    \end{aligned}
    \end{equation}
    to search for the strongest reflection angle that is not in ${\cal A}_{{\rm out},k}$, where $\widetilde{\bf h}_{k,t}^{(2)}({\bf s}_b)$ is defined by
    \begin{equation}\label{equ:channel_cascaded_app_temp_2}
    \begin{aligned}[b]
    \widetilde{\bf h}_{k,t}^{(2)}({\bf s}_b; {\cal A}_{{\rm out},k})
    =&\sum_{m,i} \sum_{j=0}^{|{\cal A}_{{\rm out},k}|-1} a_{t,m,i} d_{k,t,j} {f}_{b,i,j,k} {\bf v}_m ,
    \end{aligned}
    \end{equation}
    whose parameters $\{d_{k,t,j}\}$ are calculated by \eqref{equ:obs_ris_ue_nlos_est_LS}.

\end{itemize}

{\color{black}We summarize the estimation method in Algorithm \ref{alg:CE_Stat}. The complexity of Algorithm \ref{alg:CE_Stat} is ${\cal O}(BKTM |{\cal D}|+M I^3 J+M I^2 B K J+ I K J^3+M I B K J^2)$, which is linear in the size of the angle set $|{\cal D}|$. While a large $|{\cal D}|$ increases complexity, the computation time of Algorithm \ref{alg:CE_Stat} does not contribute to the processing delay of instantaneous channel estimation and beamforming design. Therefore, Algorithm \ref{alg:CE_Stat} can be executed with low priority when there are no urgent tasks for the CPU.
}

\section{Passive Beamforming Algorithm}\label{sec:passive_BF}
Recall that the passive beamforming problem formulation to maximize the summation of achievable rate:
\begin{subequations}
\begin{align}
    {\mathcal{P}}{(\text{A})}
  \;\;\; {\bf{s}}_t^\ast  & =\; \underset{\bf{s} }{\rm{argmax}}  \; \;
\sum_{K=1}^{K} {R}_{k,t}({\bf s}) \notag\\
 & \;\; {\text{s.t.}} \;  s_{n} \in \{1,2,\cdots,L\}, \; \forall  n, \label{equ:PA_c1}\\
 & \quad \quad \frac{1}{K} \sum_k || {\bf w}_{k,t}({\bf s})||_{\rm 2}^2 \leq P_{\rm T},
\end{align}
\end{subequations}
where the channel and the achievable rate are given by
\begin{align}
{\bf h}_{k,t}({\bf s})
=&\sum_{i,j} {\bf G}_{t}^{(i)} {\rm diag} \left({\bm \nu}(\phi_{\rm in}^{(i)},\phi_{{\rm out},k}^{(j)};{\bf s})\right) {\bf h}_{{\rm r},k,t}^{(j)}, \label{equ:channel_cascaded_opt}\\
{R}_{k,t}({\bf s})= & \log\left( 1+ \frac{\left|{\bf h}_{k,t}({\bf s})^{\rm T} {\bf w}_{k,t}({\bf s})  \right|^2 }
{\sum_{i \neq k} \left|{\bf h}_{k,t}({\bf s})^{\rm T} {\bf w}_{i,t}({\bf s})  \right|^2+\sigma_{\rm z}^2} \right). \label{equ:downlink_rate2}
\end{align}
In this section, we propose low-complexity algorithm to address the discrete constraint in \eqref{equ:PA_c1}.

\subsection{Continuous Approximation for Discrete Variables}
We propose tractable approximation to address the discrete constraint on $s_{n}$ in ${\mathcal{P}}{(\text{A})}$ by a three-step method.

\subsubsection{Reparameterize $s_{n,t}$ by One-Hot Vectors}
We introduce a one-hot vector $\overline{\bf s}_{n}=[\overline{s}_{n,1},\overline{s}_{n,2},\cdots,\overline{s}_{n,L}]^{\rm T}$ for the state of the $n$-th RIS element,  where $\overline{s}_{n,\ell} \in \{0,1\}$ and $\sum_\ell \overline{s}_{n,\ell}=1$. Then, the RIS response of the $n$-th element for the $i$-th incident ray and $j$-th reflection ray can be equivalently denoted by a function of $\overline{\bf s}_n$:
\begin{equation}\label{equ:channel_RIS_resposne2}
\begin{aligned}[b]
\nu(\phi_{\rm in}^{(i)},\phi_{\rm out}^{(j)};s_{n})=\sum_{\ell=1}^L
\overline{s}_{n,\ell} \nu(\phi_{\rm in}^{(i)},\phi_{\rm out}^{(j)};\ell).
\end{aligned}
\end{equation}

Letting $\overline{\bf S}=[\overline{\bf s}_1,\overline{\bf s}_2,\cdots,\overline{\bf s}_N]$ and substituting \eqref{equ:channel_RIS_resposne2} into \eqref{equ:channel_cascaded_opt}, the channel ${\bf h}_{k,t}({\bf s})$ is equivalently represented  by ${\bf h}_{k,t}(\overline{\bf S})$.  Consequently, the rate and precoder become ${R}_{k,t}(\overline{\bf S})$ and ${\bf w}_{k,t}(\overline{\bf S})$, respectively. Then, ${\mathcal{P}}{(\text{A})}$ is equivalently written by
\begin{subequations}
\begin{align}
    {\mathcal{P}}{(\text{A-1})}
  \; & \max_{  \overline{\bf S}
 } \; \;
 \sum_{K=1}^{K} {R}_{k,t}(\overline{\bf S}) \notag\\
 & {\text{s.t.}} \;   \sum_\ell \overline{s}_{n,\ell}=1, \; \forall  n,\label{equ:constraint_overline_s_sum}\\
  & \quad\;\;\; \overline{s}_{n,\ell} \in \{0,1\}, \; \forall  \ell, n, \label{equ:discrete_2}\\
  & \quad\; \frac{1}{K} \sum_k || {\bf w}_{k,t}(\overline{\bf S})||_{\rm 2}^2 \leq P_{\rm T}. \label{equ:power_constrant2}
\end{align}
\end{subequations}

\subsubsection{Continuous Relaxation with Sparse Promoting Constraint}
However, the binary constraint \eqref{equ:discrete_2} is still challenging. To address this issue, we relax the constraints in \eqref{equ:constraint_overline_s_sum} and \eqref{equ:discrete_2} as follows:
\begin{subequations}
\begin{align}
&\sum_\ell \sqrt{\overline{s}_{n,\ell}}=1, \; \forall  n,\label{equ:constraint_overline_s_sum_3}\\
  &  0\leq \overline{s}_{n,\ell} \leq 1, \; \forall  \ell, n. \label{equ:discrete_3}
\end{align}
\end{subequations}
The key intuition is that we prefer large $\overline{s}_{n,\ell}$ to enhance the RIS response according to \eqref{equ:channel_RIS_resposne2}. However, $\overline{s}_{n,\ell} \leq \sqrt{\overline{s}_{n,\ell}}$ for $0\leq \overline{s}_{n,\ell} \leq 1$, with equality achieved only when $\overline{s}_{n,\ell} \in \{0,1\}$. Therefore, constraint \eqref{equ:constraint_overline_s_sum_3} promotes the sparsity of $\overline{\bf s}_{n}$, encouraging $\overline{s}_{n,\ell}$ to take values close to $0$ or $1$. Thus we have
\begin{subequations}
\begin{align}
    {\mathcal{P}}{(\text{A-2})}
  \; & \max_{  \overline{\bf S}
 } \; \;
 \sum_{K=1}^{K} {R}_{k,t}(\overline{\bf S}) \notag\\
 & \; {\text{s.t.}} \;\; \eqref{equ:power_constrant2}, \eqref{equ:constraint_overline_s_sum_3}, \eqref{equ:discrete_3}. \notag
\end{align}
\end{subequations}

\subsubsection{Equivalent Transformation to Unconstrained Problem}
In practice, an unconstrained problem is preferred, as it can be easily tackled using stochastic gradient descent (SGD)-based algorithms, which are well-implemented in PyTorch functions \cite{paszke2017automatic}.  To this end, we introducing auxiliary variables $\widetilde{\bf S}=[\widetilde{\bf s}_1,\widetilde{\bf s}_2,\cdots,\widetilde{\bf s}_N]$ with  $\widetilde{\bf s}_n=[\widetilde{s}_{n,1},\widetilde{s}_{n,2},\cdots,\widetilde{s}_{n,L}]^{\rm T} \in {\mathbb R}^L$. Then, $\overline{\bf s}_n$ is represented by $\widetilde{\bf s}_n$ using softmax function:
\begin{equation}\label{equ:s_constraint_repar}
\begin{aligned}[b]
\overline{\bf s}_n=\left({\rm Softmax}(\widetilde{\bf s}_n)\right)^2,
\end{aligned}
\end{equation}
which has incorporated constraints \eqref{equ:constraint_overline_s_sum_3} and \eqref{equ:discrete_3} implicitly.
In addition, we set $\gamma(\widetilde{\bf S})$ in \eqref{equ:precoder_rzf} as follows for the RZF precoder:
\begin{equation}\label{equ:power_zf_2}
\begin{aligned}[b]
\gamma(\widetilde{\bf S})= \frac{  K  P_{\rm T}} {  \left\| {\bf H}_t^\star(\widetilde{\bf S}) \left({\bf H}_t^{\rm T}(\widetilde{\bf S}) {\bf H}_t^\star(\widetilde{\bf S}) + \varsigma {\bf I}_K  \right)^{-1} \right\|_F^2},
\end{aligned}
\end{equation}
to incorporate the power constraint in \eqref{equ:power_constrant2}.
Finally, we have an unconstrained optimization problem
\begin{subequations}
\begin{align}
    {\mathcal{P}}{(\text{A-3})}
  \; & \max_{  \widetilde{\bf S}
 } \; \; \sum_{K=1}^{K} {R}_{k,t}(\widetilde{\bf S})
 \notag
\end{align}
\end{subequations}
We solve ${\mathcal{P}}{(\text{A-3})}$ by utilizing Adam optimizer \cite{Adam} in PyTorch with learning rate $0.1$.

\section{Simulation}
\subsection{Simulation Setup}
\begin{figure}
[!t]
\centering
\includegraphics[width=0.80\columnwidth]{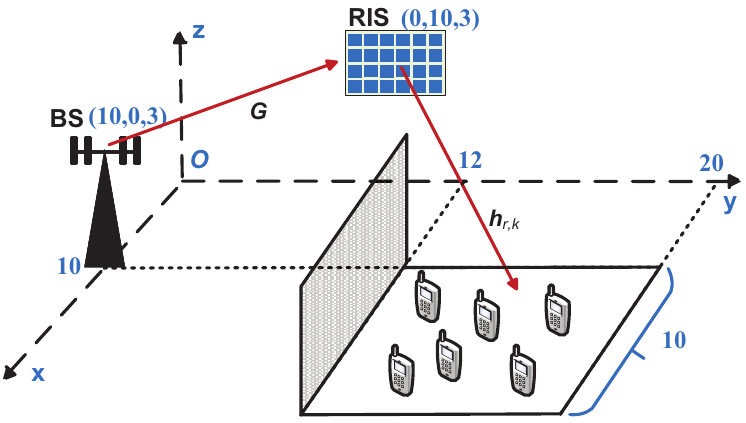}
\caption{The simulated RIS-aided $K$-user MISO communication scenario comprising of one $M$-antenna BS and one $N$-element IRS.}
\label{indoor_2user}
\end{figure}

We examine an indoor femtocell network operating at 2.4 GHz with a bandwidth of 60 kHz, as shown in {\figurename~\ref{indoor_2user}}. Users are randomly distributed in a 10 m × 8 m area, served by one BS and one RIS. {\color{black}Channel coefficients are generated using the 3GPP ray-tracing model \cite[Section 7.5]{38901} with parameters for the Indoor-Office scenario \cite[Table 7.5-6]{38901}, and the LoS channel condition is set with a K-factor of 3 dB.} The RIS is a $8 \times 8$ uniform planar array with $N=64$ elements. Each element experiences an average reflection loss of approximately 1.5 dB, primarily due to the RF switches. This performance is 3 dB better than that of a reflective antenna of the same size, as it also benefits from structural scattering. {\color{black}To evaluate the advantages of the pixel RIS, we investigate two setups varying the sizes of the BS array and the number of users:
\begin{itemize}
\item {\bf Setup 1}: BS uses a $4 \times 1$ ULA to serve $K=2$ users for spatial multiplexing.
\item {\bf Setup 2}: BS uses a $4 \times 2$ UPA to serve $K=4$ users for spatial multiplexing.
\end{itemize}
}

For channel estimation, we set $T=20$ and $B=16$. We generate 100 snapshots with random locations, and 40 random channel realizations are generated to verify the channel estimation accuracy and the final achievable rate. We set $P_{\rm T}=\sigma_{\rm z}^2=\sigma_{\rm u}^2=1$, and the average signal-to-noise ratio (SNR) for wireless propagation is defined by:
\begin{equation}
\begin{aligned}[b]
{\rm{SNR}}\triangleq{\mathbb E}_{t,k} \left[ \frac{\|{\bf G}_t\|_{\rm F}^2 \|{\bf h}_{{\rm r},k,t}\|_2^2 }{MN^2} \right]
.
\end{aligned}
\end{equation}

\begin{figure}
[!t]
\centering
\includegraphics[width=.95\columnwidth]{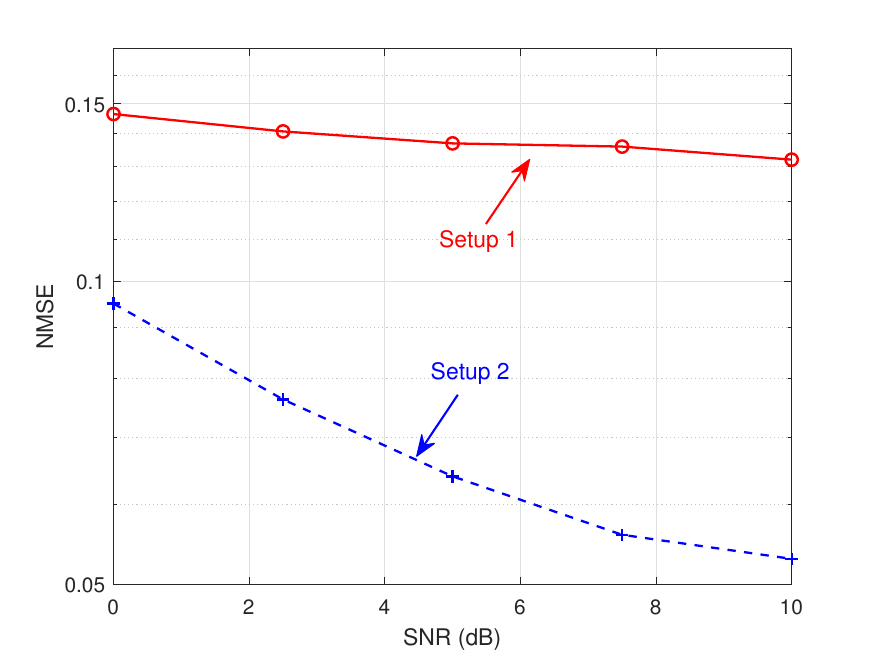}
\caption{SNR versus NMSE for different setups.}
\label{snr_vs_NMSE}
\end{figure}

{\color{black}To evaluate the effectiveness of the proposed solution, we consider the following baselines:
\begin{itemize}
\item {\bf Random}: Each element selects a random state. This baseline does not require channel estimation for RIS configuration.
\item {\bf Best Sounding}: This scheme selects the best state among the $B=16$ RIS state configurations during channel sounding.
\item {\bf Group-Opt \cite{greedyoptStatisticCE2023WSA,greedyopt2019TWC}}: In this scheme, all the elements select the same state, and the optimal state is selected by exhaustive search given perfect CSI.
\item {\bf Phased Array}: This scheme employs traditional RIS with the same size, with phase shifters optimized to maximize the sum rate under perfect CSI assumption, while the BS employs the RZF precoder shown in \eqref{equ:precoder_rzf}.
\end{itemize}
}

\subsection{Simulation Results}
We first evaluate the effectiveness of the proposed channel estimation algorithm in {\figurename~\ref{snr_vs_NMSE}}. The estimation NMSE for each wireless realization is assessed using 100 random RIS configurations not observed during the channel sounding phases. The proposed algorithm performs well in both setups, achieving an NMSE below 0.15, with notably better performance in setup 2. This improvement is due to the increased number of BS antennas $M$ and users $K$, which enhances the estimation of dominated rays, as discussed in Section \ref{sec:CE_inst}.

\begin{figure}
[!t]
\centering
\includegraphics[width=.95\columnwidth]{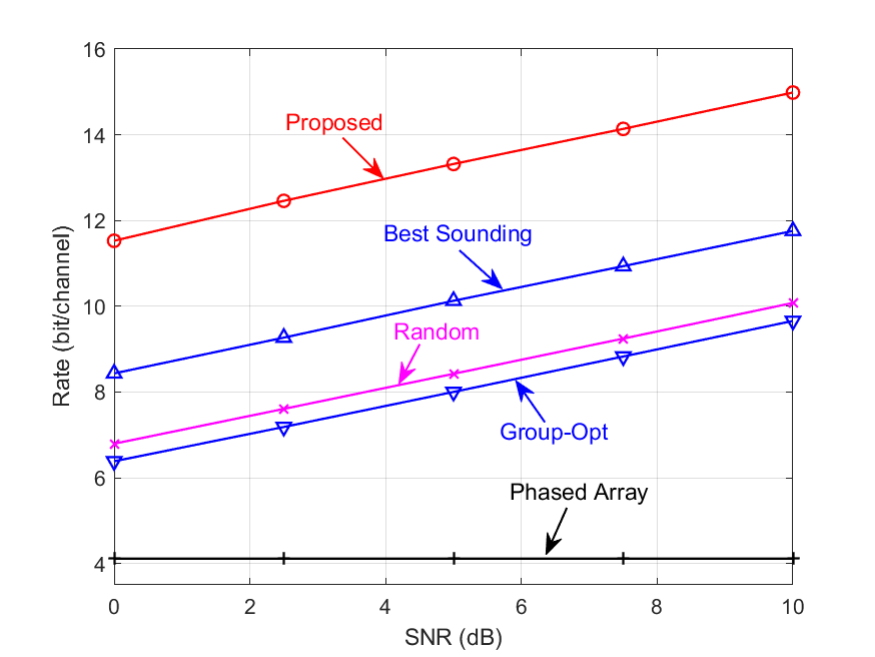}
\caption{SNR versus achievable rate for downlink MU-MISO under setup 1 ($M=4$, $N=64$, and $K=2$).}
\label{R_vs_SNR}
\end{figure}

\begin{figure}
[!t]
\centering
\includegraphics[width=.95\columnwidth]{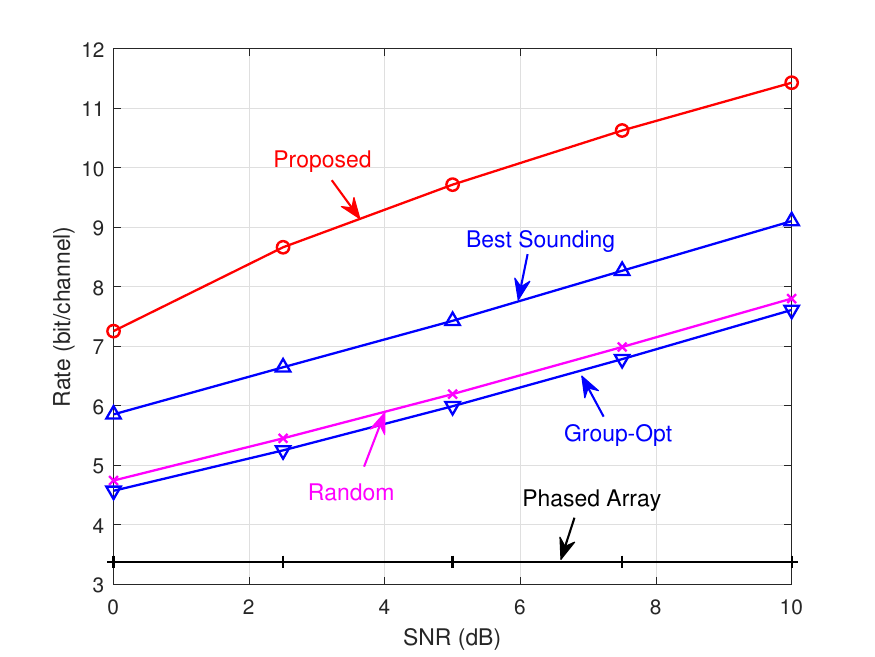}
\caption{SNR versus achievable rate for downlink MU-MISO under setup 2 ($M=8$, $N=64$, and $K=4$).}
\label{R_vs_SNR_2}
\end{figure}

{\color{black}Next, we compare the achievable rates of the proposed solution with baselines in setup 1 ({\figurename~\ref{R_vs_SNR}}) and setup 2 ({\figurename~\ref{R_vs_SNR_2}}). The phased array performs poorly due to the strong BS-RIS LoS link and high channel correlation among multiple users, leading to interference as a key bottleneck. Consequently, the achievable rate shows minimal improvement with increasing SNR. This is also evidenced by the decrease in achievable rate with more users for spatial multiplexing, despite a proportional increase in BS antennas, as shown in the comparison of the two figures. In contrast, the pixel RIS outperforms the phased array even with random configurations, as its elements can reflect an incident ray in two focused directions (as shown in {\figurename~\ref{resp_state_1}}), which potentially enhances the multiuser channel condition. Furthermore, our proposed solution achieves significant performance gains in all cases, with an SNR improvement of 8 dB in setup 1 and 6 dB in setup 2 compared to the best baseline.}

\section{Conclusion}\label{conclusion}
In this paper, we proposed a passive beamforming and channel estimation solution for pixel-based RIS with a non-separable state response. We introduced a DNN to model the RIS response function while ensuring low computational and memory requirements. We then presented a tractable channel estimation algorithm that focuses on the channel responses of dominated scattering rays, with statistical knowledge estimated from historical observations. Finally, we implemented a low-complexity passive beamforming algorithm using a novel continuous approximation to configure the discrete RIS states. Simulations showed that the proposed solution achieved significant gains over various baselines.

\bibliographystyle{IEEEtran}
\bibliography{IEEEabrv, bibliography}

\end{document}